\documentclass{revtex4}
\usepackage{amssymb}
\usepackage{amsmath}
\usepackage{graphicx}
\usepackage{dcolumn}
\usepackage{bm}
\usepackage{amsfonts}

\begin{document}

\title{\textbf{Extensivity of Irreversible Current and Stability in Causal
Dissipative Hydrodynamics}}
\author{G. S.~Denicol, T.~Kodama, T.~Koide and Ph.~Mota}

\begin{abstract}
We extended our formulation of causal dissipative hydrodynamics [T. Koide 
\textit{et al.}, Phys. Rev. \textbf{C75}, 034909 (2007)] to be applicable to
the ultra-relativistic regime by considering the extensiveness of
irreversible currents. The new equation has a non-linear term which
suppresses the effect of viscosity. We found that such a term is necessary
to guarantee the positive definiteness of the inertia term and stabilize
numerical calculations in ultra-relativistic initial conditions. Because of
the suppression of the viscosity, the behavior of the fluid is more close to
that of the ideal fluid. Our result is essentially same as that from the
extended irreversible thermodynamics, but is different from the
Israel-Stewart theory. A possible origin of the difference is discussed.
\end{abstract}

\pacs{47.10.-g,25.75.-q}
\maketitle

\section{Introduction}

It is widely believed that the basic features of collective motion in
relativistic heavy-ion collisions can be well described by using the
(almost) ideal hydrodynamic model \cite{GeneralHydro}. Several studies on
the effects of viscosity are available in various works and seem to support
such a vision \cite%
{1st-a,1st-b,muro,2nd,heinzson,heinzson2,dkkm1,dkkm2,dkkm3}. However, the
inclusion of the dissipation into the relativistic hydrodynamics is not
trivial because the naive generalization of the Navier-Stokes equation gives
rise to the problems of acausality and instability. Various theories have
been proposed to incorporate dissipation consistent with causality and
stability: the divergence type theory \cite{muller}, the Israel-Stewart
theory \cite{II}, the extended irreversible thermodynamics \cite{Jou},
Carter's theory \cite{carter}, \"{O}ttinger-Grmela formulation \cite{OG},
the approach base on conformal field theory \cite{baier} and so on. However,
the relations between these theories are not well clarified and the
formulation of the relativistic hydrodynamics itself has not yet established.

One crucial point of these theories is that dissipative fluids behave as a
kind of non-Newtonian fluids in a relativistic energy regime \cite{tk1,tk2}.
Recently, we proposed another formulation of the relativistic dissipative
hydrodynamics, stressing this aspect \cite{dkkm1,dkkm2}. The definition of
the irreversible current is modified by the introduction of a memory
function, and we showed that such an effect is enough to solve the problem
of acausality. We also found that the theory is stable against the linear
perturbations around the hydrostatic state \cite{dkkm3}. This vision offers
new possibilities in adapting the techniques developed for \textit{Newtonian
fluids} to a \textit{non-Newtonian} regime \cite{dkkm2,tk1,tk2}.

Although our equations of the dissipative hydrodynamics can be reduced to
the truncated Israel-Stewart (IS) theory, this formulation can be
generalized for the application to ultra-relativistic situations. One of the
important assumptions in the derivation of hydrodynamics is the local
equilibrium ansatz: at any space point, there should exist an \textit{finite}
extension of the fluid which is described by the thermodynamic laws in
equilibrium. We refer such an element of the fluid to a fluid cell. The
irreversible currents are phenomenologically defined so that the second law
of thermodynamics is satisfied for each fluid cell. These fluid cells permit
superpositions and need not to be exclusive, each other, but they have to be
of finite size to apply thermodynamics. However, in the usual hydrodynamic
formulation, the finiteness of the fluid cell is considered to be irrelevant
and the irreversible currents are defined by applying the second law for 
\textit{densities} of extensive quantities. This does not give rise to
problems in the (relativistic) Navier-Stokes theory, because thermodynamic
forces instantaneously produce irreversible currents. That is, the time
scale of deformation of a fluid cell is considered infinitely large in
comparison with this scale. This is also the case when the relaxation time,
which characterizes the memory effect, is small compared to the variation
scale of the fluid. However, when the relaxation time is of the same order
as the variation scale, we have to distinguish the quantities affected by
the change in internal degrees of freedom from the global kinematic degrees
of freedom in introducing memory effects.

In this paper, we rederive the causal dissipative hydrodynamics by taking
the finite size of the fluid cell into account. For simplicity, we consider
the 1+1 dimensional system. We found that the equation of the bulk viscosity
has a nonlinear term. Because of the nonlinear term, the effect of bulk
viscosity is suppressed and the behavior of the fluid is closer to that of
the ideal fluid. More importantly, we found that this effect is
indispensable to implement stable numerical calculations in
ultra-relativistic initial conditions.

This paper is organized as follows. In Sec.\ref{chap:2}, we discuss the new
formulation of relativistic dissipative hydrodynamics by taking the finite
size of the fluid cell into account. The restriction for the parameters
which are consistent with causality is discussed in Sec.\ref{chap:3}. In Sec.%
\ref{chap:4}, we apply our theory to the 1+1 dimensional scaling solution of
the Bjorken model. In Sec.\ref{chap:5}, we implements numerical simulations.
We apply the smoothed particle formulation to solve the hydrodynamics
numerically. Here, we discussed three examples, the shock formation, the
expansion to vacuum with Landau initial condition and the appearance of
nonperiodic oscillations similar to turbulence. We show that the result of
our formulation can be justified from the viewpoint of the extended
irreversible thermodynamics in Sec.\ref{chap:6}. The relation between our
theory and the Israel-Stewart theory is discussed in Sec.\ref{chap:7}. Sec.%
\ref{chap:8} is devoted to concluding remarks.

\section{Extensive Measure for the Irreversible Current}

\label{chap:2}

For simplicity, we consider the case of vanishing baryon chemical potential
for the simple 1+1 dimensional system. In this case, the hydrodynamic
equations of motion can be written as only the conservation of the
energy-momentum tensor, 
\begin{equation}
\partial _{\mu }T^{\mu \nu }=0,  \label{Tmunu}
\end{equation}%
together with the thermodynamic relations. We adopt the local equilibrium
ansatz in the energy local rest frame, as proposed by Landau-Lifshitz \cite%
{LL}. The energy-momentum tensor is given by 
\begin{equation}
T^{\mu \nu }=\left( \varepsilon +P+\Pi \right) u^{\mu }u^{\nu }-\left( P+\Pi
\right) g^{\mu \nu },
\end{equation}%
where $\varepsilon $, $P$, $u^{\mu }$ and $\Pi $ are, respectively, the
energy density, pressure, four-velocity and bulk viscosity. From Eq.(\ref%
{Tmunu}), we obtain the entropy production rate in terms of densities, 
\begin{equation}
\partial _{\mu }\left( su^{\mu }\right) =-\frac{1}{T}\Pi \partial _{\mu
}u^{\mu }.  \label{dsmu/dt}
\end{equation}

We should remember that the local thermal equilibrium ansatz must be applied
to a fluid cell which, in principle, has a finite volume defined by the
coarse-grained size of microscopic degrees of freedom. The thermodynamic
laws should be applied to the integrated quantities of the fluid inside each
cell. To see this more clearly, let us introduce the volume $V^{\ast }$ of
such a fluid cell. That is, $V^{\ast }$ is the volume of the fluid at the
point $\vec{r}$, inside of which the fluid is considered to be homogeneous
and satisfies the thermodynamic laws in equilibrium. The flow of the fluid
deforms such a cell so that its volume is a function of time. If we follow
the fluid flow given by the velocity field $\vec{v}$, the time variation of $%
V^{\ast }$ is given by 
\begin{equation}
\frac{1}{V^{\ast }}\frac{dV^{\ast }}{dt}=\nabla \cdot \vec{v}
\end{equation}%
or in a covariant form, 
\begin{equation}
\partial _{\mu }\left( \sigma u^{\mu }\right) =0,  \label{continuity}
\end{equation}%
where we have introduced the proper reference density $\sigma$ by 
\begin{equation}
\sigma =\frac{1}{V}=\frac{\gamma }{V^{\ast }},
\end{equation}%
where $\gamma$ is the Lorentz factor.

Now, let us denote the extensive measure of the entropy inside this volume
as $\tilde{s}=sV=s/\sigma$. Then we rewrite Eq.(\ref{dsmu/dt}) as%
\begin{equation}
T\frac{d\tilde{s}}{d\tau } = -\tilde{J} F =-\tilde{\Pi} \partial _{\mu
}u^{\mu },  \label{eqn:epuc}
\end{equation}%
with 
\begin{equation}
\tilde{\Pi}=\Pi V=\frac{\Pi }{\sigma }
\end{equation}%
is the extensive measure inside the fluid cell of the irreversible current $%
\Pi$.

One can see that Eq.(\ref{eqn:epuc}) has the structure that the net entropy
production in the cell is given by the product of the irreversible
displacement $\tilde{\Pi}$ occurred in the cell and the corresponding
thermodynamic force field $F=\partial _{\mu }u^{\mu}$ in the cell. It should
be noted that, from Eq. (\ref{continuity}), the thermodynamic force is
reexpressed as 
\begin{equation}
F = \partial _{\mu }u^{\mu} = \sigma \frac{d}{d\tau} \left( \frac{1}{\sigma}
\right).
\end{equation}
This result means that the thermodynamic force for the fluid cell is given
by the change of the cell volume, indicating clearly the physical meaning of
the bulk viscosity: the resistance to the change of the volume of the system.

In the relativistic Navier-Stokes theory (the Landau-Lifshitz theory), it is
assumed that the bulk viscosity per volume element is produced by the
thermodynamic force without any retardation, 
\begin{equation}
\tilde{\Pi}=-\tilde{\eta}F=-\frac{\zeta }{\sigma }\partial _{\mu }u^{\mu },
\end{equation}%
where $\tilde{\eta}$ represents a extensive measure of coupling (total
charge) for the whole matter inside the cell. We thus identify $\zeta$ as
the bulk viscosity coefficient. By multiplying $\sigma$ for both sides, we
reproduce the usual result of the Landau-Lifshitz theory \cite{LL}. That is,
the finite size effect does not affect the definition of irreversible
currents in the Navier-Stokes theory. However, it is by now well known that
the derived equation has the problem of acausality and instability \cite%
{dkkm1,dkkm2,dkkm3}. To solve these difficulties, we introduce a memory
effect to the irreversible current by using a memory function.

When microscopic and macroscopic scales are clearly separated, the time
scale of the variation of the fluid cell itself is infinitely large compared
to the microscopic scales, and only the change in the internal degrees of
freedom is relevant. If, however, this is not true, they are entangled and
we have to distinguish their roles. One is the change of the internal
degrees of freedom with the transient effects (memory effect) and the other
the motion of the global kinematic degrees of freedom (fluid cell
deformation). As the simplest memory function which can be reduced to the
differential equation, we apply 
\begin{equation}
G\left( \tau ,\tau ^{\prime }\right) =\frac{1}{\tau _{\mathrm{R}}\left( \tau
^{\prime }\right) }\exp \left( -\int_{\tau ^{\prime }}^{\tau }\frac{d\tau
^{\prime \prime }}{\tau _{\mathrm{R}}\left( \tau ^{\prime \prime }\right) }%
\right) .  \label{Kernel}
\end{equation}%
Here, $\tau _{\mathrm{R}}$ is a relaxation time which characterizes the time
scale of the retardation. In the previous work \cite{dkkm1}, we have applied
this memory function to the thermodynamic force $\eta F$ (not $\tilde{\eta}F$%
) to introduce the retardation. As a consequence, the final form of the
derived equation for $\Pi$ is same as that of the so-called truncated form
of the IS equation, where nonlinear terms are ignored. However as we
mentioned before, when we deal with memory effects, we should not use
densities. This is because memory effects relate different thermodynamic
states of the matter for different times and this depends on the size of the
system not necessarily in an extensive manner. Then the memory effect should
be applied to the integrated quantity of a fluid cell.

In this paper, we introduce the memory effect of the thermodynamic force
field on an extensive measure for the irreversible current. Then the bulk
viscosity consistent with causality is given by 
\begin{equation}
\tilde{\Pi}\left( \tau \right) =-\int_{\tau _{0}}^{\tau }d\tau ^{\prime
}G\left( \tau ,\tau ^{\prime }\right) \frac{\zeta }{\sigma }\partial _{\mu
}u^{\mu }+\tilde{\Pi}_{0}\exp \left( -\int_{\tau _{0}}^{\tau }\frac{d\tau
^{\prime }}{\tau _{\mathrm{R}}\left( \tau ^{\prime }\right) }\right) ,
\label{Pi_tilda}
\end{equation}%
where $\Pi _{0}$ is the initial value given at $\tau _{0}$. This integral
form is equivalent to the solution of the following differential equation, 
\begin{equation}
\tau _{\mathrm{R}}\frac{d}{d\tau }\tilde{\Pi}+\tilde{\Pi}=-\frac{\zeta }{%
\sigma }\partial _{\mu }u^{\mu }.  \label{eqn:neweqbulk1}
\end{equation}%
This equation can be reexpressed in terms of the density as follows, 
\begin{equation}
\tau _{\mathrm{R}}\frac{d}{d\tau }\Pi +\Pi =-\left( \zeta +\tau _{\mathrm{R}%
}\Pi \right) \partial _{\mu }u^{\mu }.  \label{eqn:neweqbulk2}
\end{equation}%
On the other hand, the corresponding equation in our previous paper \cite%
{dkkm1} where the memory effect is directly applied to the densities has the
form, 
\begin{equation}
\tau _{\mathrm{R}}\frac{d}{d\tau }\Pi +\Pi =-\zeta \partial _{\mu }u^{\mu }.
\label{trunc}
\end{equation}%
The difference of Eq.(\ref{eqn:neweqbulk2}) from Eq. (\ref{trunc}) is the
presence of the term $-\tau _{\mathrm{R}}\Pi \partial _{\mu }u^{\mu }$.

It should be noted that Eq. (\ref{eqn:neweqbulk2}) also does not explicitly
depend on the cell volume $1/\sigma$. This is because the size of the cell
volume is irrelevant as far as the length is much larger than the mean-free
path and much smaller than the typical hydrodynamic scale. For later
convenience, we call the causal dissipative hydrodynamics without the finite
size effect, Eq.(\ref{trunc}) as \textquotedblleft linear causal dissipative
hydrodynamics (LCDH)\textquotedblright, whereas the one with finite size
effect, Eq. (\ref{eqn:neweqbulk2}) as \textquotedblleft nonlinear causal
dissipative hydrodynamics (NLCDH)\textquotedblright.

An important effect due to the nonlinear term is the lower bound of the bulk
viscosity. In LCDH, the bulk viscosity can, in principle, take any negative
value. In NLCDH, when the bulk viscosity becomes negatively large, the
effective bulk viscosity coefficient $\zeta_{\text{eff}}=\zeta +\tau_{%
\mathrm{R}}\Pi$ eventually changes sign and the bulk viscosity start to
increase. Thus, the bulk viscosity in NLCDH cannot be smaller than%
\begin{equation}
\Pi_{\mathrm{min}}=-\frac{\zeta}{\tau_{\mathrm{R}}},
\end{equation}
when the initial value of $\Pi$ is larger than $\Pi_{\text{min}}$. This
aspect plays an important role for the stability of numerical simulations of
ultra-relativistic cases, as we will see Sec. \ref{chap:5}.

\section{Propagation speed of sound}

\label{chap:3}

We parametrize the bulk viscosity coefficient and the relaxation time as
follows, 
\begin{align}
\zeta & =as,  \label{eqn:bulk_para} \\
\tau_{\mathrm{R}} & =\frac{\zeta}{\varepsilon+P}b,  \label{eqn:relax_para}
\end{align}
where $a$ and $b$ are arbitrary constants.

As was pointed out in \cite{dkkm1,dkkm2,dkkm3}, LCDH can be acausal
depending on the choice of the parameters. To see the limitation of the
parameters, we have to calculate the propagation speed of NLCDH. Following 
\cite{dkkm1,dkkm2,dkkm3}, we discuss the linear perturbation around the
hydrostatic state. Then the nonlinear term of the equation of the bulk
viscosity (\ref{eqn:neweqbulk2}) disappears and the dispersion relation is
same as that of LCDH \cite{dkkm2,dkkm3}. Thus, by assuming the group
velocity gives the propagating speed of the dissipative fluid, we found 
\begin{equation}
v_{\mathrm{c}}=\sqrt{1/b+\alpha},
\end{equation}
where $\alpha=\partial P/\partial\varepsilon$. To satisfy causality $v_{%
\mathrm{c}}\leq1$, the parameter $b$ should be less than $1/\left(
1-\alpha\right) $. This is completely the same restriction as the case of
LCDH \cite{dkkm2,dkkm3}.

\section{Scaling solution}

\label{chap:4}

\begin{figure}[ptb]
\includegraphics[scale=0.3]{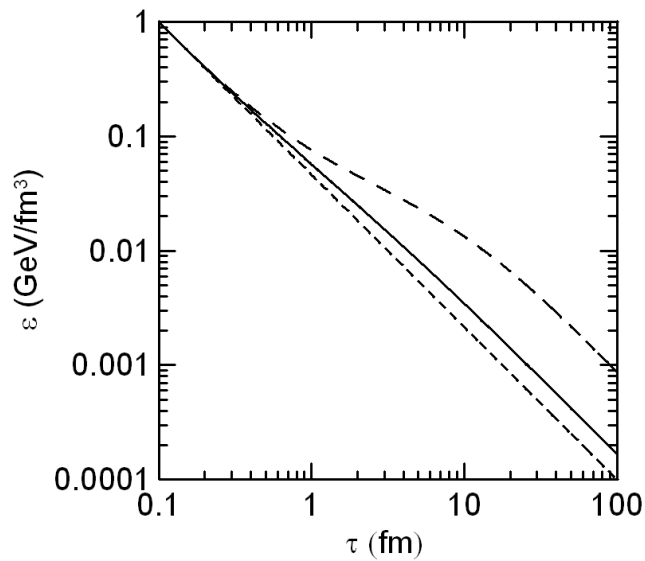}
\caption{The time evolution of the energy density of the scaling solution.
The dotted, dashed and solid lines corresponds to the ideal hydrodynamics,
LCDH and NLCDH, respectively.}
\label{fig:scaling}
\end{figure}

\begin{figure}[ptb]
\includegraphics[scale=0.3]{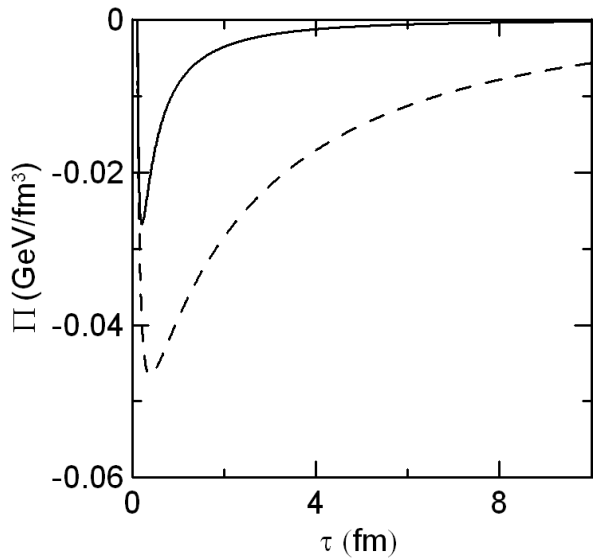}
\caption{The time evolution of the bulk viscosity of the scaling solution.
The dashed and solid lines corresponds to LCDH and NLCDH, respectively. }
\label{fig:scaling2}
\end{figure}

We apply NLCDH to the one dimensional scaling solution of the Bjorken model.
Then the hydrodynamic equations are given by 
\begin{align}
& \partial_{\tau}\varepsilon+\frac{\varepsilon+P+\Pi}{\tau}=0, \\
& \tau_{\mathrm{R}}\partial_{\tau}\Pi+\Pi=-\frac{\zeta+\tau_{\mathrm{R}}\Pi 
}{\tau},
\end{align}
where $t=\tau\cosh y$ and $x=\tau\sinh y$.

We adopt the massless ideal gas equation of state where $\alpha=1/3$. To
satisfy causality, the parameter $b$ should be larger than $3/2$. In this
calculation, we choose $b=6$. For the initial condition, we set $\varepsilon
\left( \tau_{0}\right) =1~\mathrm{GeV/fm}^{3}$ and $\Pi\left(
\tau_{0}\right) =0$ at the initial proper time $\tau_{0}=0.1$ fm.

In Fig. \ref{fig:scaling}, we show the energy density $\varepsilon$ as a
function of the proper time $\tau$. The dotted, dashed and solid lines
correspond to the ideal hydrodynamics, LCDH and NLCDH, respectively. Because
of the memory effect, the behaviors of LCDH and NLCDH are similar to that of
the ideal fluid at the early stage of the time evolution. After the time
larger than the relaxation time, the behaviors of LCDH deviates from that of
the ideal hydrodynamics. On the other hand, the behavior of NLCDH stays
close to that of the ideal hydrodynamics, that is, the effect of the bulk
viscosity is suppressed in NLCDH compared to LCDH. This is directly observed
from the behavior of the bulk viscosity as is shown in Fig. \ref%
{fig:scaling2}.

It should be noted that the scaling solution of the bulk viscosity including
the similar nonlinear term was already discussed in \cite{muro}, where the
equation derived by Israel and Stewart is discussed. The relation of our
equation and the Israel-Stewart equation will be discussed in Sec. \ref%
{chap:7}.

\section{Numerical simulations}

\label{chap:5}

\subsection{Smoothed particle formulation}

\label{chap:numeri}

To solve numerically the hydrodynamic equations, we use the Smoothed
Particle Hydrodynamic (SPH) method. The original idea of the SPH method is
to obtain an approximate solution of hydrodynamics by parameterizing the
fluid into a set of effective particles \cite{SPH}. Because of its
flexibility to adapt to complex geometries, the SPH method has also been
extensively applied to the relativistic heavy ion reactions to perfom an
event-by-event analysis of the data \cite{ref:and}. For the sake of
convenience, we reproduce below the ideas and basic equations of the SPH
approach shown in \cite{dkkm2}.

Let us consider a distribution $a\left( \mathbf{r},t\right) $ of any
extensive physical quantity, $A$. In a system like the hot and dense matter
created in heavy ion collisions, the behavior of $a\left( \mathbf{r}%
,t\right) $ contains the effects of whole microscopic degrees of freedom. We
are not interested in the extremely short wavelength behavior of $a\left( 
\mathbf{r},t\right) $ but rather in global behaviors which are related
directly to the experimental observables. Therefore, we would like to
introduce a coarse-graining procedure for $a$. To do this, we introduce the
kernel function $W\left( \mathbf{r}-\tilde{\mathbf{r}},h\right) $ which maps
the original distribution $a$ to a coarse-grained version $a_{\mathrm{CG}}$
as, 
\begin{equation}
a_{\mathrm{CG}}\left( \mathbf{r},t\right) =\int a\left( \tilde{\mathbf{r}}%
,t\right) W\left( \mathbf{r}-\tilde{\mathbf{r}},h\right) d\tilde {\mathbf{r}}
\label{as}
\end{equation}
where $W$ is normalized, 
\begin{equation}
\int W\left( \tilde{\mathbf{r}},h\right) d\tilde{\mathbf{r}}=1,
\end{equation}
and has a bounded support of the scale of $h$, 
\begin{equation}
W\left( \mathbf{r},h\right) \rightarrow0,\ \ \left\vert \mathbf{r}%
\right\vert \gtrsim h,
\end{equation}
satisfying 
\begin{equation}
\underset{h\rightarrow0}{\lim}W\left( \tilde{\mathbf{r}},h\right)
=\delta\left( \tilde{\mathbf{r}}\right) .
\end{equation}
Here, $h$ is a typical length scale for the coarse-graining in the sense
that the kernel function $W$ introduces a cut-off in short wavelength of the
order of $h$. Thus we will take this value as the scale of coarse graining
in the QCD dynamics (i.e., the mean-free path of partons) to obtain the
hydrodynamics of QGP ($h\simeq0.1~\mathrm{fm}$).

The second step is to approximate this coarse-grained distribution $a_{%
\mathrm{CG}}\left( \mathbf{r},t\right) $ by replacing the integral in Eq.(%
\ref{as}) by a summation over a finite and discrete set of points, $\left\{ 
\mathbf{r}_{\alpha}(t),\alpha=1,\dots,N_{\mathrm{SPH}}\right\}$, 
\begin{equation}
a_{\mathrm{SPH}}\left( \mathbf{r},t\right) =\sum_{\alpha=1}^{N_{\mathrm{SPH}%
}}A_{\alpha}\left( t\right) W\left( \left| \mathbf{r}-\mathbf{r}_{\alpha
}(t)\right| \right) .  \label{aSPH}
\end{equation}
If the choice of $\left\{ A_{\alpha}(t),\alpha=1,\dots,N_{\mathrm{SPH}%
}\right\} $ and $\left\{ \mathbf{r}_{\alpha}(t),\alpha=1,\dots ,N_{\mathrm{%
SPH}}\right\} $ are appropriate, the above expression should converge to the
coarse-grained distribution $a_{\mathrm{CG}}$ for large $N_{\mathrm{SPH}}$.
Parameters $\left\{ A_{\alpha}(t),\alpha=1,\dots ,N_{\mathrm{SPH}}\right\} $
and $\left\{ \mathbf{r}_{\alpha}(t),\alpha =1,\dots,N_{\mathrm{SPH}}\right\} 
$ should be determined from the dynamics of the system. In practice, we
first choose the reference density $\sigma^{\ast}$ which is conserved,%
\begin{equation}
\frac{\partial\sigma^{\ast}}{\partial t}+\nabla\cdot\mathbf{j}=0,
\end{equation}
where $\mathbf{j}$ is the current associated with the density $%
\sigma^{\ast}. $ Then, we note that the following ansatzes, 
\begin{align}
\sigma_{\mathrm{SPH}}^{\ast}\left( \mathbf{r},t\right) & =\sum_{\alpha
=1}^{N_{\mathrm{SPH}}}\nu_{\alpha}W\left( |\mathbf{r}-\mathbf{r}_{\alpha
}(t)|\right) , \\
\mathbf{j}_{\mathrm{SPH}}\left( \mathbf{r},t\right) & =\sum_{\alpha =1}^{N_{%
\mathrm{SPH}}}\nu_{\alpha}\frac{d\mathbf{r}_{\alpha}(t)}{dt}W\left( |\mathbf{%
r}-\mathbf{r}_{\alpha}(t)|\right) ,
\end{align}
satisfies the equation, 
\begin{equation}
\frac{\partial\sigma_{\mathrm{SPH}}^{\ast}}{\partial t}+\nabla\cdot \mathbf{j%
}_{\mathrm{SPH}}=0,
\end{equation}
where $\nu_{\alpha}$' s are constant. By using the normalization of $W$, we
have 
\begin{equation}
\int_{\mathrm{SPH}}\sigma^{\ast}\left( \mathbf{r},t\right) d^{3}\mathbf{r=}%
\sum_{\alpha=1}^{N_{\mathrm{SPH}}}\nu_{\alpha}.
\end{equation}
Then we can interpret the quantity $\nu_{\alpha}$ as the conserved quantity
attached at the point $\mathbf{r}=\mathbf{r}_{\alpha}(t)$. Therefore, the
distribution $\sigma_{\mathrm{SPH}}^{\ast}\left( \mathbf{r},t\right) $ is a
sum of small piece-wise distribution, carrying the density,%
\begin{equation}
\nu_{\alpha}W\left( |\mathbf{r}-\mathbf{r}_{\alpha}(t)|\right) .
\end{equation}
These pieces are referred to as "SPH-particles".

Using the above reference density and the extensive nature of $A,$ we can
write $A_{\alpha}$ in Eq.(\ref{aSPH}) as%
\begin{align}
A_{\alpha}\left( t\right) =\nu_{\alpha}\frac{a\left( \mathbf{r}_{\alpha
},t\right) }{\sigma^{\ast}\left( \mathbf{r}_{\alpha},t\right) }
\end{align}
which represents the quantity $A$ carried by the SPH particle at the
position $\mathbf{r}=\mathbf{r}_{\alpha}(t)$. In fact, the total amount of $%
A $ of the system at the instant $t$ is given by%
\begin{align}
A\left( t\right) =\sum_{\alpha=1}^{N_{\mathrm{SPH}}}A_{\alpha}\left(
t\right) .
\end{align}

In the ideal fluid, the entropy density is chosen as the reference density
and the dynamics of the parameters $\left\{ \mathbf{r}_{\alpha}(t),\alpha
=1,\dots,N_{\mathrm{SPH}}\right\} $ are determined from the variational
principle from the action of ideal hydrodynamics. The entropy density is,
however, not conserved for the dissipative fluid. Thus we introduce a new
conserved quantity, the specific proper density $\sigma$, which is defined
by the flow of the fluid, 
\begin{align}
\partial_{\mu}\left( \sigma u^{\mu}\right) =0,  \label{conserv-sigmav}
\end{align}
and we will use it as the reference density for viscous fluids. Here, the
four-velocity $u^{\mu}$ is defined in terms of the local rest frame of the
energy flow (Landau frame). The specific density is expressed in the SPH
form as 
\begin{align}
\sigma^{\ast}\left( \mathbf{r},t\right) =\sum_{\alpha=1}^{N_{\mathrm{SPH}%
}}\nu_{\alpha}W\left( |\mathbf{r}-\mathbf{r}_{\alpha}(t)|\right) ,
\end{align}
where $\sigma^{\ast}=\sigma u^{0}$ is the specific density in the laboratory
frame and $\nu_{\alpha}$ is the inverse of the specific volume of the SPH
particle $\alpha$. In this work, the specific volume should be interpreted
as the volume of the fluid cell, and hence $\nu_{\alpha}$ is the inverse of
the cell volume. However, as we showed, the finial results do not depend on
this choice and we set $\nu_{\alpha}=1$ for simplicity. As for the kernel $%
W\left( \mathbf{r}\right) $, we use the spline function.

It should be mentioned that this procedure is only possible provided that
the lines of flow in space defined by the velocity field $u^{\mu}$ do not
cross each other during the evolution in time. That is, if there appear
turbulence or singularities in the flow lines, the above definition of
Lagrange coordinates can fail.

Now we apply this method to NLCDH in $1+1$ dimension. We have to solve the
evolution equation of the bulk viscosity in the SPH scheme. For this, we
express the viscosity as 
\begin{equation}
\Pi=\sum_{\alpha=1}^{N_{\mathrm{SPH}}}\nu_{\alpha}\left( \frac{\tilde{\Pi}}{%
\gamma}\right) _{\alpha}W\left( |\mathbf{r}-\mathbf{r}_{\alpha}(t)|\right) ,
\end{equation}
The time evolution of the term $\tilde{\Pi}_{\alpha}$ can be calculated as%
\begin{equation}
\frac{d\tilde{\Pi}_{\alpha}}{dt}=-\frac{\zeta}{\sigma_{\alpha}^{\ast}\tau_{%
\mathrm{R}}}\left( \partial_{\mu}u^{\mu}\right) _{\alpha}-\frac {1}{%
\gamma_{\alpha}\tau_{\mathrm{R}}}\tilde{\Pi}_{\alpha}  \label{BulkSPH}
\end{equation}
where $\gamma_{\alpha}$ is the Lorentz factor of the $\alpha$-th particle.
In the following, we denote the quantity in the observable frame with the
asterisk. It should be noted that we solve Eq. (\ref{eqn:neweqbulk1})
instead of (\ref{eqn:neweqbulk2}) in the numerical calculations.

At the same time, using the SPH expression for the entropy density $s^{\ast}$
in the observable frame, 
\begin{equation}
s^{\ast}=\sum_{\alpha=1}^{N_{\mathrm{SPH}}}\nu_{\alpha}\left( \frac{s}{\sigma%
}\right) _{\alpha}W\left( |\mathbf{r}-\mathbf{r}_{\alpha}(t)|\right) ,
\end{equation}
and the evolution of the entropy per SPH particle is given by\textbf{\ } 
\begin{equation}
\frac{d}{dt}\left( \frac{s}{\sigma}\right) _{\alpha}=-\frac{1}{T}\frac {%
\Pi_{\alpha}}{\sigma_{\alpha}^{\ast}}\left( \partial_{\mu}u^{\mu}\right)
_{\alpha}.
\end{equation}
where $s=s^{\ast}/u^{0}$ is the proper entropy density. In the above
expressions, the relaxation time $\tau_{\mathrm{R}},$ viscosity coefficient $%
\zeta$ and temperature $T$ are functions of space and time, so that they
should be evaluated at the position of each particle $\alpha.$

Finally, we need to express the momentum conservation equation by the SPH
variables. We write the space component of energy-momentum equation of
continuity in terms of the reference density, 
\begin{equation}
\sigma\frac{d}{d\tau}\left( \frac{\epsilon+P+\Pi}{\sigma}u^{i}\right)
+\partial_{i}\left( P+\Pi\right) =0.  \label{Motion}
\end{equation}

It should be noted that there exist ambiguities within the resolution of the
coarse-graining size $h$ to express the equation of motion in the SPH form.
However, in the ideal fluid, the SPH equation of motion can be derived by
the variational method uniquely. Thus, we obtain the equation of motion by
using the same SPH parametrization to Eq.(\ref{Motion}), 
\begin{align}
& \sigma_{\alpha}\frac{d}{d\tau_{\alpha}}\left( \frac{\epsilon_{\alpha
}+P_{\alpha}+\Pi_{\alpha}}{\sigma_{\alpha}}u_{\alpha}^{i}\right) =  \notag \\
& \sum_{\beta=1}^{N_{\mathrm{SPH}}}\nu_{\beta}\sigma_{\alpha}^{\ast} \left( 
\frac{P_{\beta}+\tilde{\Pi}_{\beta}\sigma_{\beta}}{\left( \sigma_{\beta}^{*}
\right) ^{2}} +\frac{P_{\alpha}+\tilde{\Pi}_{\alpha}\sigma_{\alpha}}{\left(
\sigma_{\alpha}^{*} \right) ^{2}} \right) \partial_{i}W\left( |\mathbf{r}%
_{\alpha}-\mathbf{r}_{\beta}(t)|\right) ,  \label{MotionSPH}
\end{align}
where the right hand side of Eq.(\ref{MotionSPH}) corresponds to the term $%
\partial_{i}\left( P+\Pi\right) $ written in terms of the SPH
parametrization. We remark that in the case of vanishing viscosity our
result is reduced to the expression derived with variational principle for
ideal fluids.

By separating the acceleration and force terms in Eq.(\ref{MotionSPH}), we
obtain our final expression of the equation of motion for each SPH particle, 
\begin{equation}
\mathbf{M}\frac{d\mathbf{u}}{dt}=\mathbf{F},
\end{equation}
where the mass matrix $\mathbf{M}$ and the force term $\mathbf{F}$ are
defined as%
\begin{align}
M_{ij} & =\gamma\left( \epsilon+P+\Pi\right) \delta_{ij}+Au_{i}u_{j},
\label{Mij} \\
F_{j} & =-\partial_{j}\left( P+\Pi\right) +Bu_{j},  \label{Fj}
\end{align}
with 
\begin{align}
A & =-\frac{1}{\gamma}\left[ \alpha(\varepsilon+P+\Pi)+\frac{\zeta}{\tau_{%
\mathrm{R}}}+\Pi\right] ,  \label{A} \\
B & =A\frac{\gamma^{2}}{\sigma^{\ast}}\frac{d\sigma^{\ast}}{dt}+\frac{\Pi }{%
\tau_{\mathrm{R}}},  \label{B}
\end{align}
where $\alpha=\partial P/\partial\varepsilon$. It should be noted that the
expression of $A$ in NLCDH is different from that of in LCDH \cite{dkkm2},%
\begin{equation}
A_{\mathrm{LCDH}}=-\frac{1}{\gamma}\left[ \alpha\left( \varepsilon
+P+\Pi\right) +\frac{\zeta}{\tau_{\mathrm{R}}}\right] .
\end{equation}
To carry out the calculation in this scheme, the mass matrix $\mathbf{M}$
must be nonsingular. However, this is not guaranteed in LCDH. For example,
consider the 1+1 dimensional system with ultra-relativistic fluid velocity, $%
u\approx\gamma$. Then the mass in LCDH becomes 
\begin{equation}
M_{\mathrm{LCDH}}\approx\gamma\left[ \left( 1-\alpha\right) \left(
\varepsilon+P+\Pi\right) -\frac{\zeta}{\tau_{\mathrm{R}}}\right] .
\end{equation}
As we discussed the bulk viscosity in LCDH does not have a lower bound, we
see that $M_{\mathrm{LCDH}}$ can be zero, while in NLCDH, we get 
\begin{eqnarray}
M_{\mathrm{NLCDH}} & \approx\gamma & \left[ \left( 1-\alpha\right) \left(
\varepsilon+P+\Pi\right) -\Pi-\frac{\zeta}{\tau_{\mathrm{R}}}\right]
\geq\gamma\left( 1-\alpha\right) \left( \varepsilon+P-\frac{\zeta}{\tau_{%
\mathrm{R}}}\right)  \notag \\
& =& \gamma\left( 1-\alpha\right) \left( \varepsilon+P\right) \left( 1-\frac{%
1}{b}\right) >0.
\end{eqnarray}
Here, we used the expression of the bulk viscosity coefficient and the
relaxation time, Eqs. (\ref{eqn:bulk_para}) and (\ref{eqn:relax_para}). The
mass does not vanish in NLCDH and the simulation is stable even for the
ultra-relativistic situations.

\subsection{Shock formation}

\label{sec:shock}

\begin{figure}[ptb]
\begin{minipage}{.45\linewidth}
\includegraphics[scale=0.3]{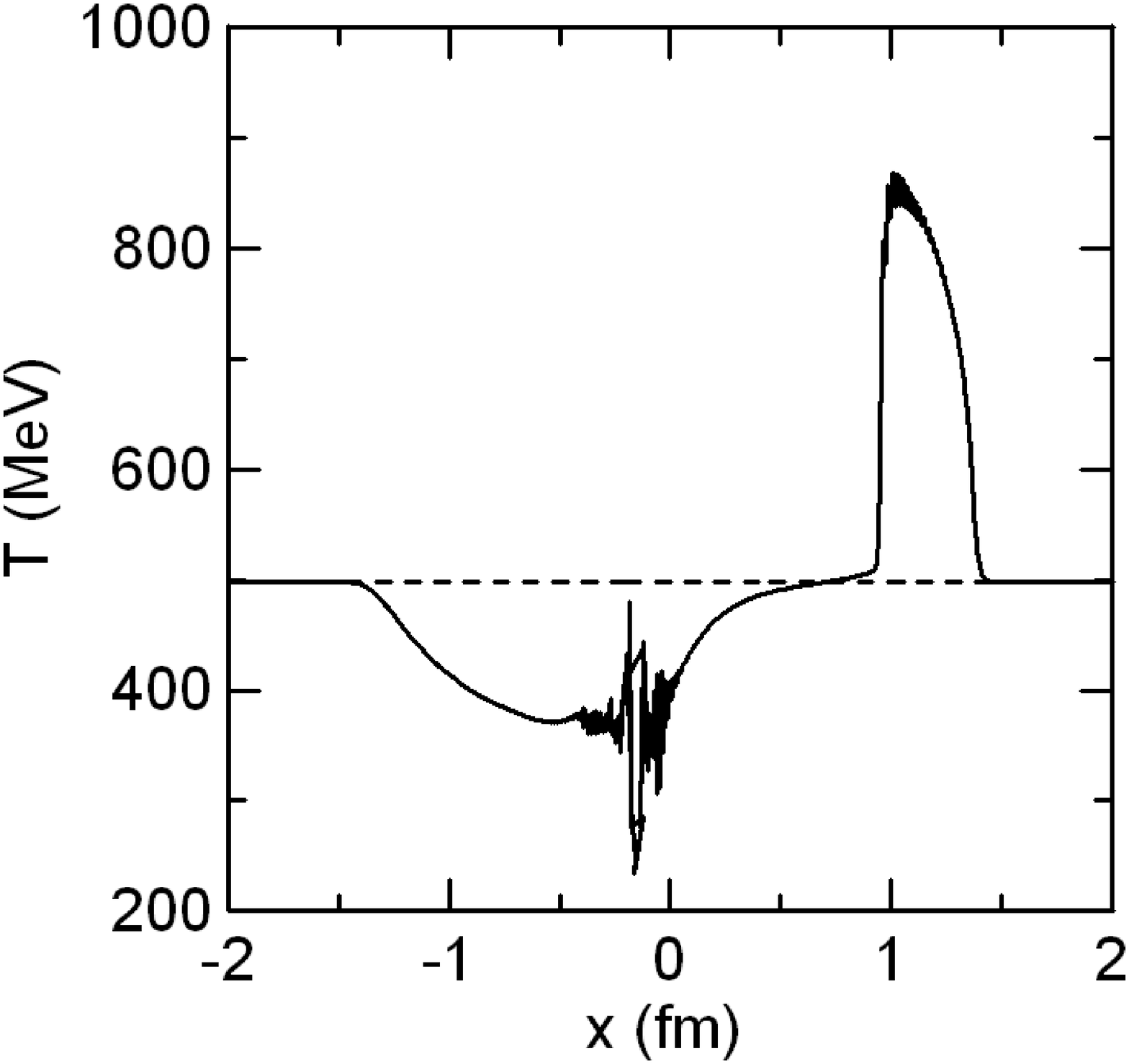}
\caption{The temperature in the shock formation calculated in LCDH with $a=1$ at $t=0.75$ fm,
starting from the homogeneous initial condition (dotted line).}
\label{temp}
\end{minipage}
\begin{minipage}{.45\linewidth}
\includegraphics[scale=0.3]{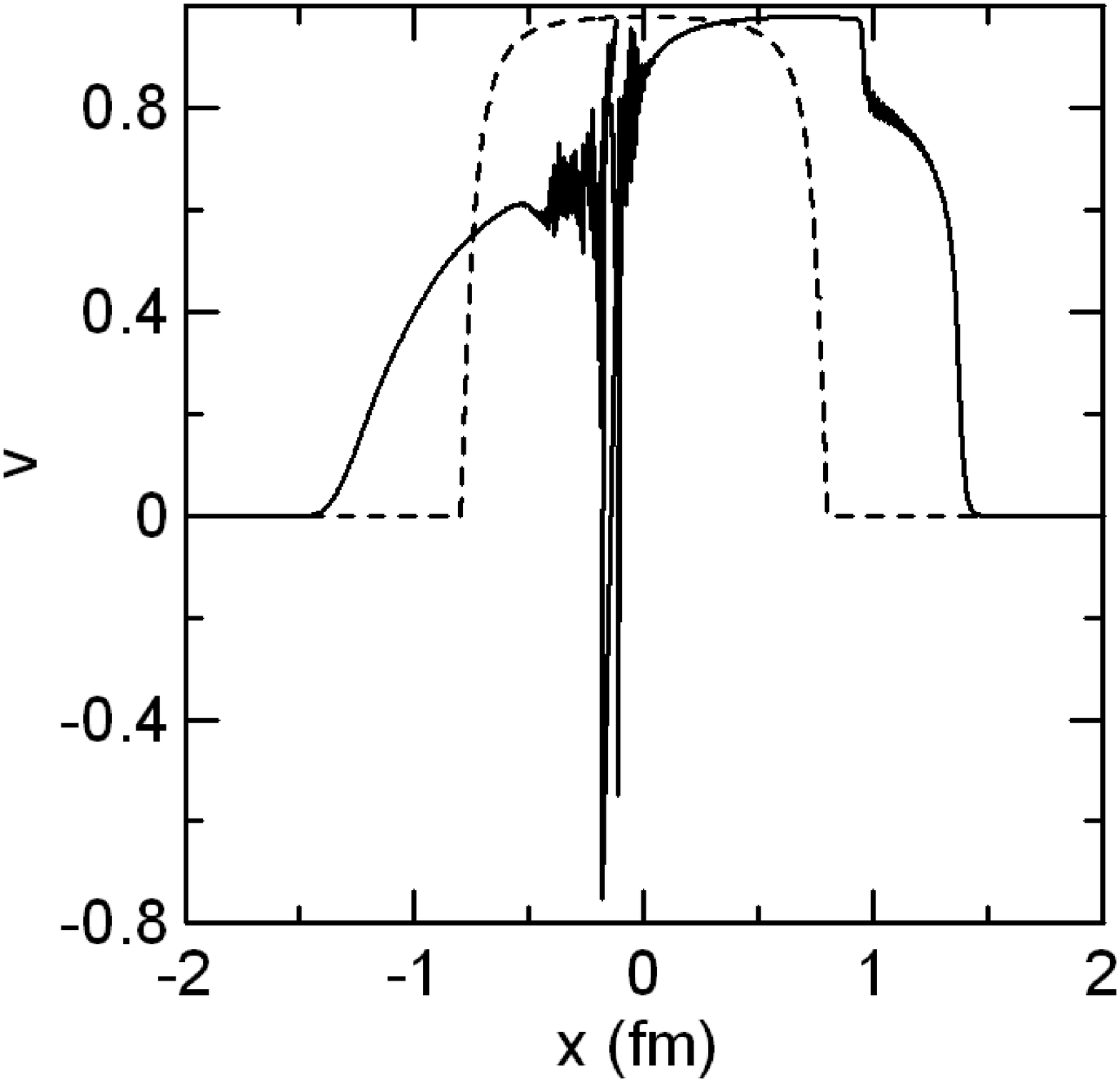}
\caption{The velocity in the shock formation calculated in LCDH with $a=1$ at $t=0.75$ fm.
The initial velocity (dotted line) at the maximum is $\gamma=5$.}
\label{vel}
\end{minipage}
\begin{minipage}{.45\linewidth}
\includegraphics[scale=0.3]{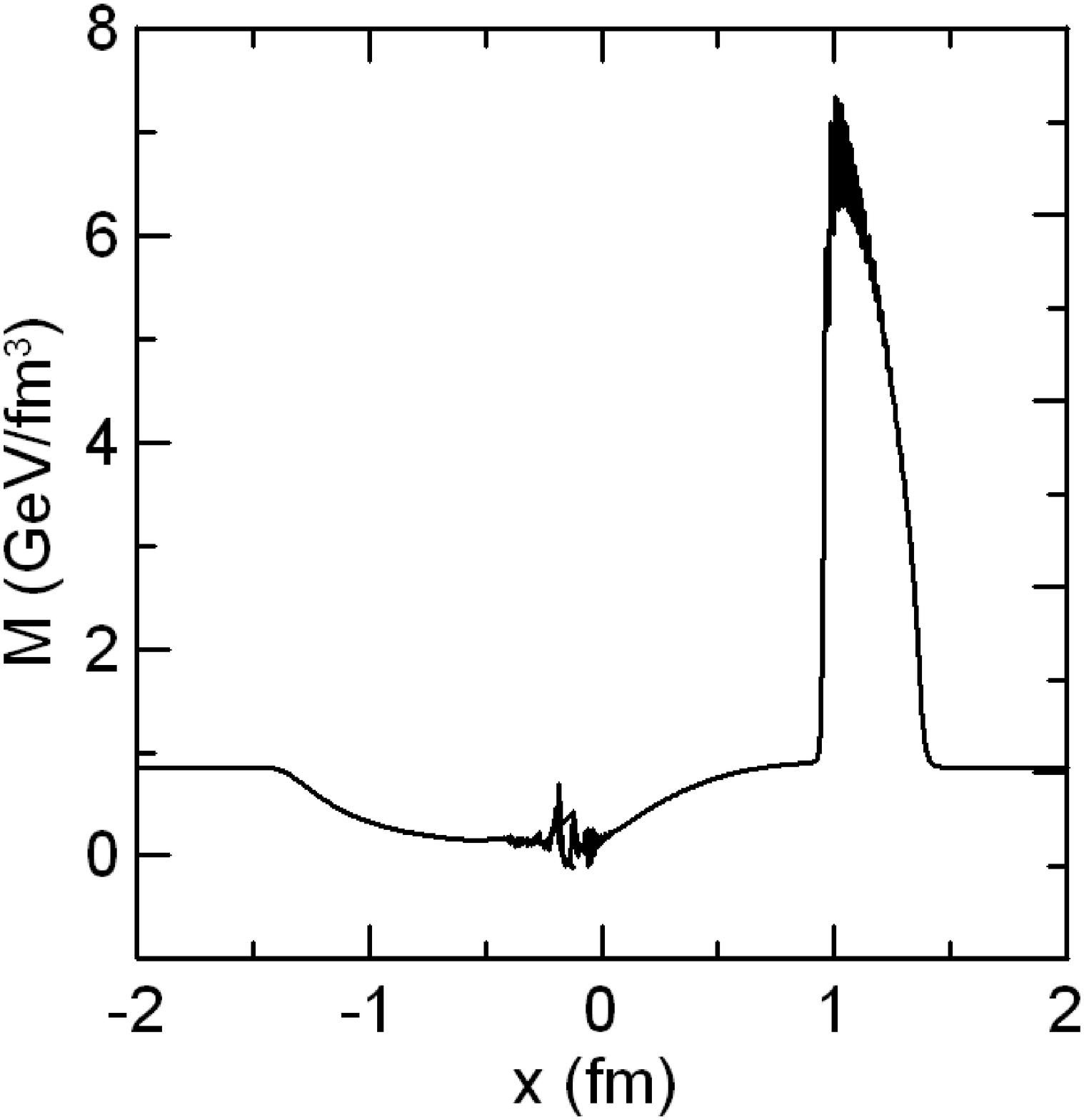}
\caption{The mass matrix in the shock formation calculated in LCDH with $a=1$ at $t=0.75$ fm.
The mass matrix crosses zero around the minimum.}
\label{mass}
\end{minipage}
\end{figure}

As was discussed in \cite{dkkm2}, in the LCDH scheme, the numerical
calculation can be carried out with the help of the additional viscosity for 
$\gamma=2$. Here, we show that this scheme becomes unstable for
ultra-relativistic initial conditions. We use the same additional viscosity
as the one proposed in \cite{dkkm2}. In Figs. \ref{temp} and \ref{vel}, we
show, respectively, the temperature and velocity profiles calculated in LCDH
with $a=1$ at $t=0.75$ fm for the initial triger velocity $\gamma=5$. The
dotted lines are for the initial condition. We can see that the calculation
becomes unstable and rapid oscillations appear around $x=0$ and the
calculation eventually collapses. This is due to the vanishing of the mass
term. The behavior of the mass term is shown in Fig. \ref{mass}. We can see
that the mass becomes zero at the point where the calculation shows the
rapid oscillation.

\begin{figure}[ptb]
\begin{minipage}{.45\linewidth}
\includegraphics[scale=0.3]{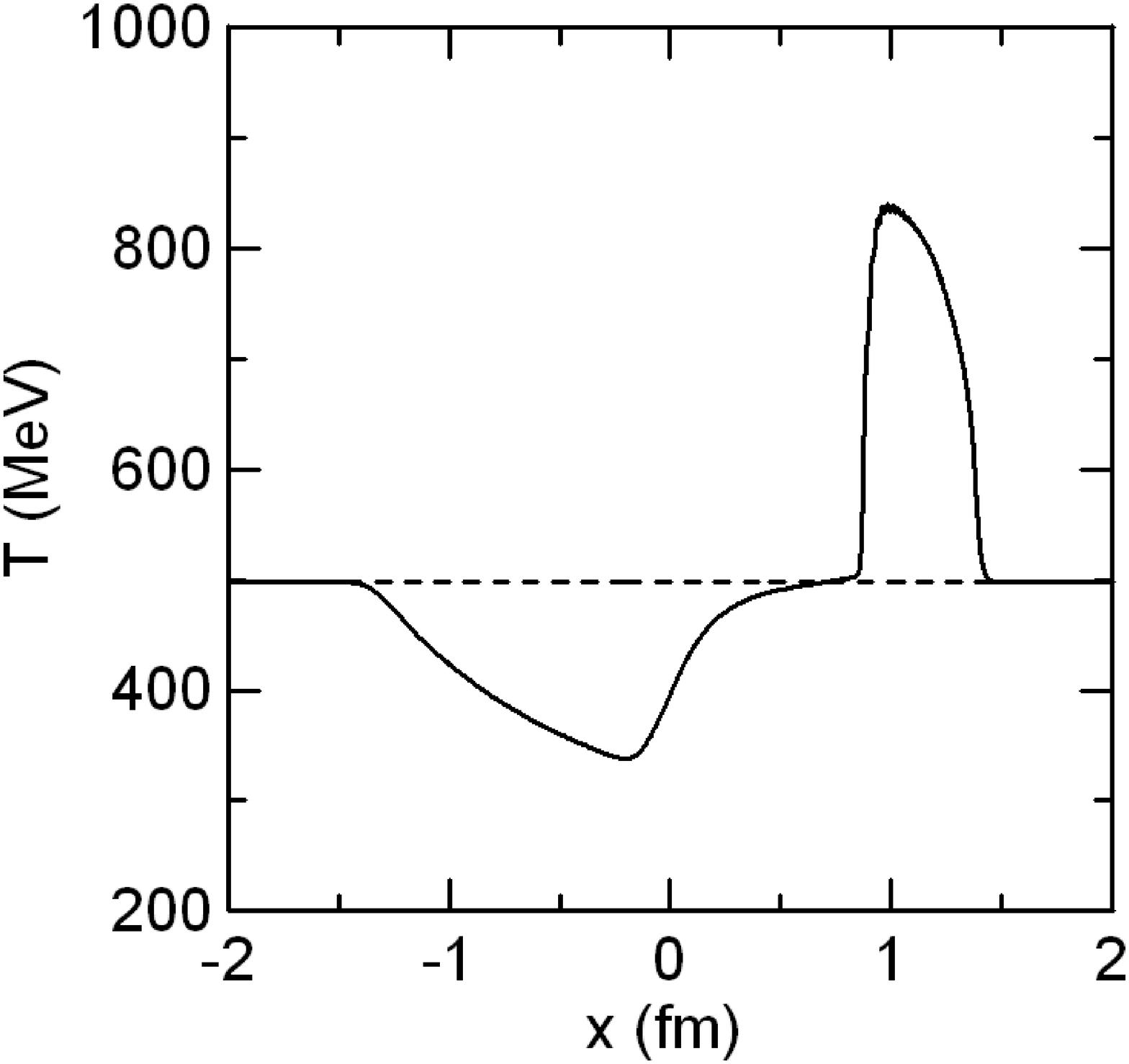}
\caption{The temperature in the shock formation calculated in NLCDH with $a=1$ and $b=6$ at $t=0.75$ fm,
starting from the homogeneous initial condition (dotted line).}
\label{temp2}
\end{minipage}
\begin{minipage}{.45\linewidth}
\includegraphics[scale=0.3]{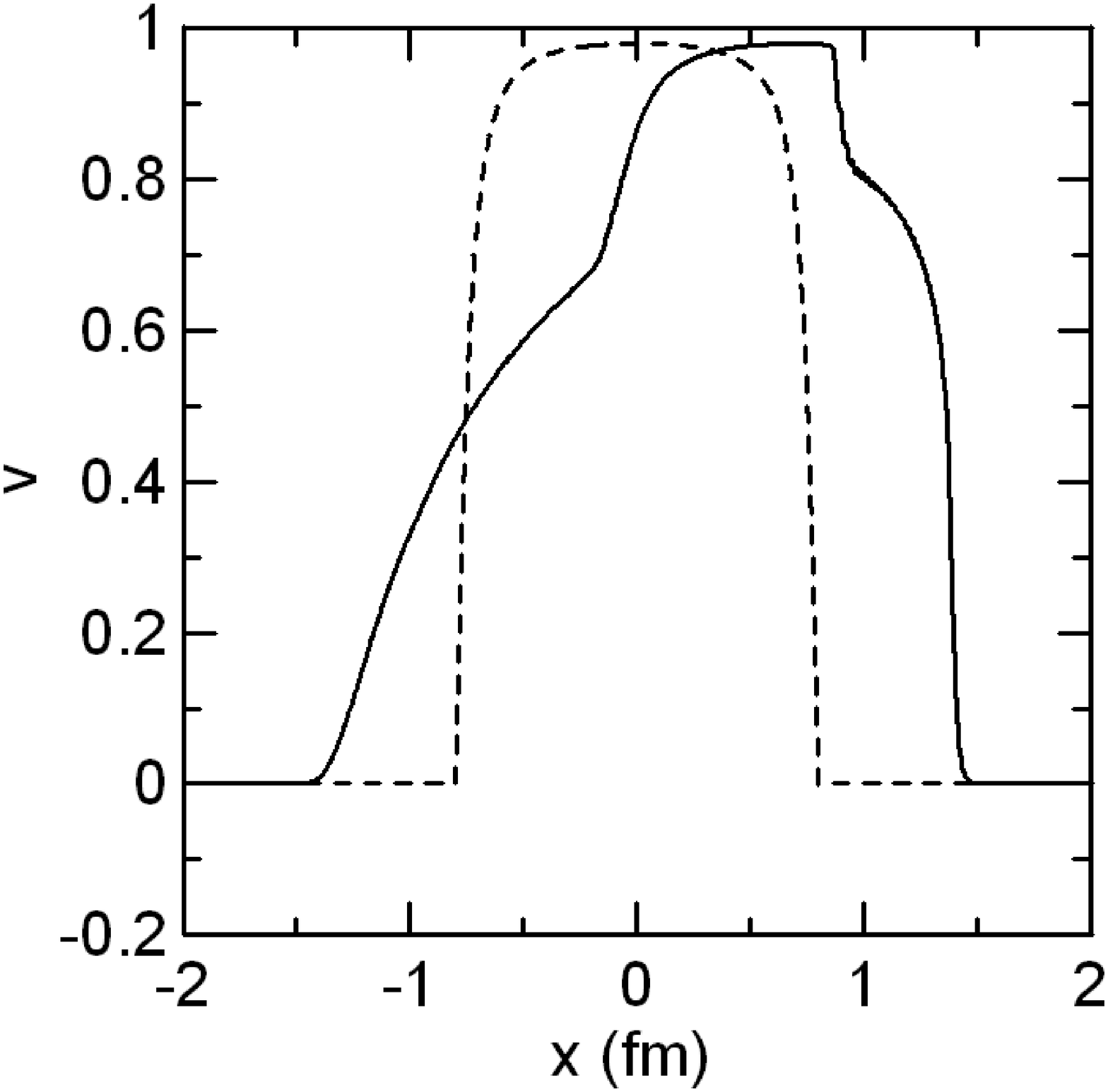}
\caption{The velocity in the shock formation calculated in NLCDH with $a=1$ and $b=6$ at $t=0.75$ fm.
The initial velocity (dotted line) at the maximum is $\gamma=5$.}
\label{vel2}
\end{minipage}
\begin{minipage}{.45\linewidth}
\includegraphics[scale=0.3]{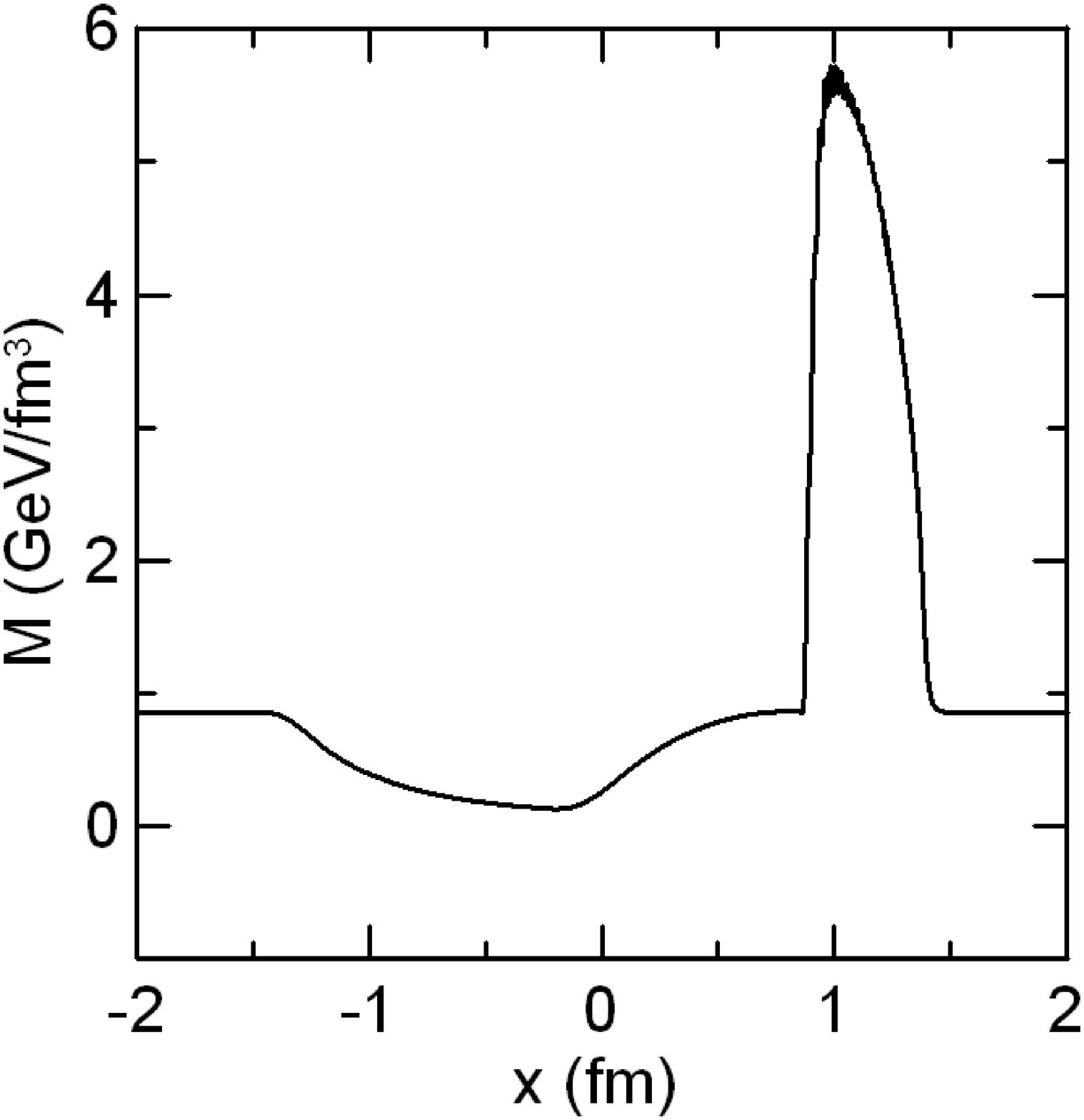}
\caption{The mass matrix in the shock formation calculated in NLCDH with $a=1$ and $b=6$ at $t=0.75$ fm.
The mass matrix does not cross zero.}
\label{mass2}
\end{minipage}
\end{figure}

On the other hand, as was shown in the previous section, the mass does not
vanish in NLCDH. Thus we can implement the numerical calculation with the
ultra-relativistic initial condition without numerical singularities. In
Figs. \ref{temp2} and \ref{vel2}, we show, respectively, the temperature and
velocity distributions calculated in NLCDH with $a=1$ and $b=6$ at $t=0.75$
fm, starting from the same initial condition. We can see that the
calculation remains stable, because the mass does not vanish as is shown in
Fig.\ref{mass2}.

In all simulations of the present work, we consistently use the additional
viscosity in the NLCDH scheme. That is, we introduce the nonlinear term for
the coarse-grain viscosity used in \cite{dkkm2}.

\subsection{Landau initial condition}

\label{sec:landau}

\begin{figure}[ptb]
\begin{minipage}{.45\linewidth}
\includegraphics[scale=0.3]{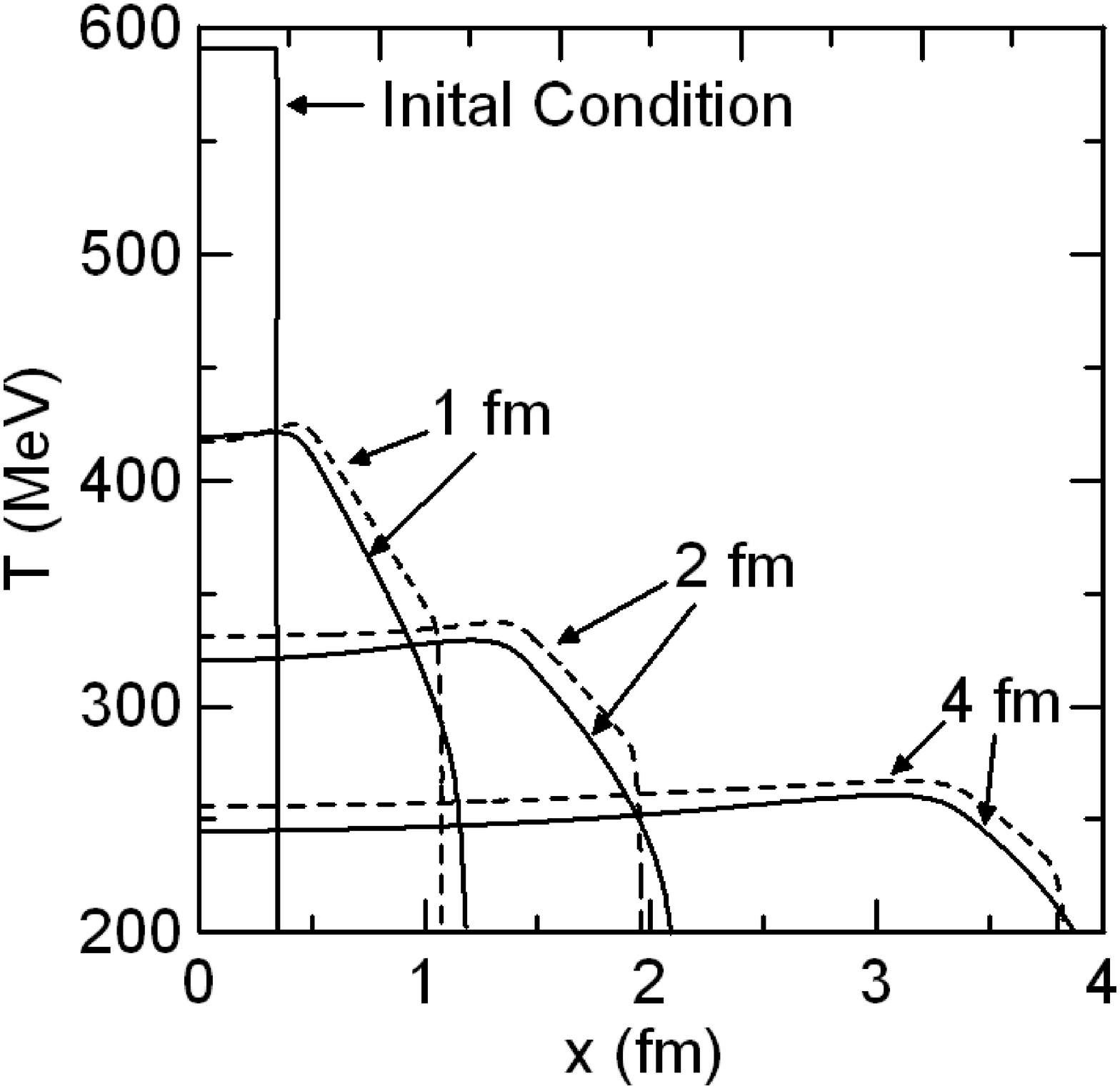}
\caption{The evolution of the temperature with the Landau initial condition using $a=0.1$ for $t=1$, $2$ and $3$ fm.
The solid and dotted lines denote the calculations in NLCDH and LCDH, respectively.}
\label{landau_temp}
\end{minipage}
\hspace{1cm} 
\begin{minipage}{.45\linewidth}
\includegraphics[scale=0.3]{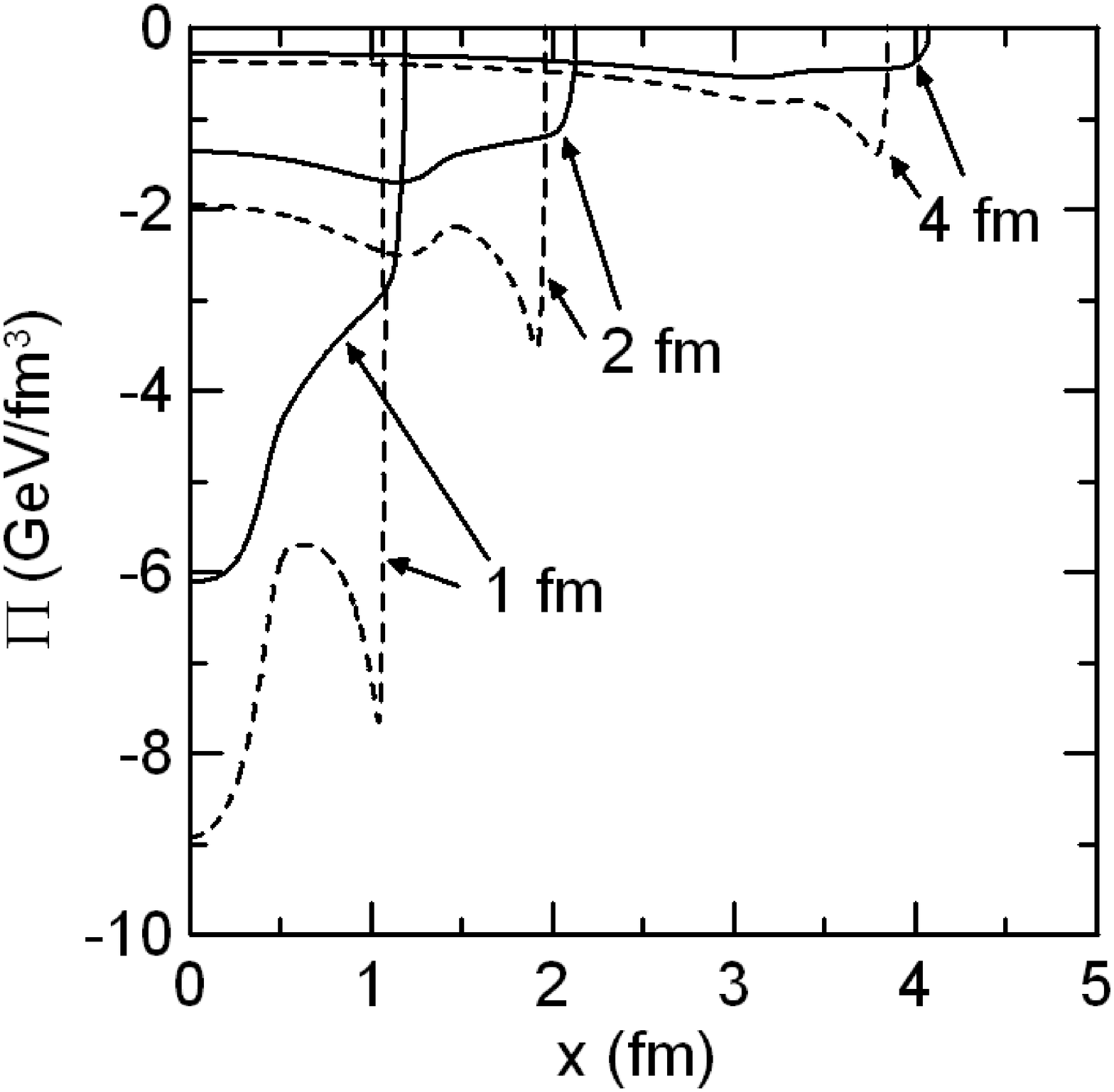}
\caption{The evolution of the bulk viscosity with the Landau initial condition using $a=0.1$ for $t=1$, $2$ and $3$ fm.
The solid and dotted lines denote the calculations in NLCDH and LCDH, respectively.}
\label{bulk}
\end{minipage}
\end{figure}

Here, we discuss the expansion of the fluid to vacuum and compare the
cooling process of LCDH and NLCDH. We use the Landau initial condition where
the initial temperature is $590$ MeV and the initial size is $0.7$ fm. In
Fig.\ref{landau_temp}, we show the evolution of the temperature with $a=0.1$
and $b=6$ for $t=1$, $2$ and $4$ fm, from the top. The solid and dotted
lines represents the results of NLCDH and LCDH, respectively. One can see
that the cooling and expansion of the fluid of the NLCDH are faster than
that of LCDH, similarly to the case of the scaling solution. This is because
the bulk viscosity in NLCDH is supressed by the nonlinear term in comparison
with LCDH. This is explicitly shown in Fig. \ref{bulk}, where the evolutions
of the bulk viscosity are plotted.

\begin{figure}[ptb]
\begin{minipage}{.45\linewidth}
\includegraphics[scale=0.3]{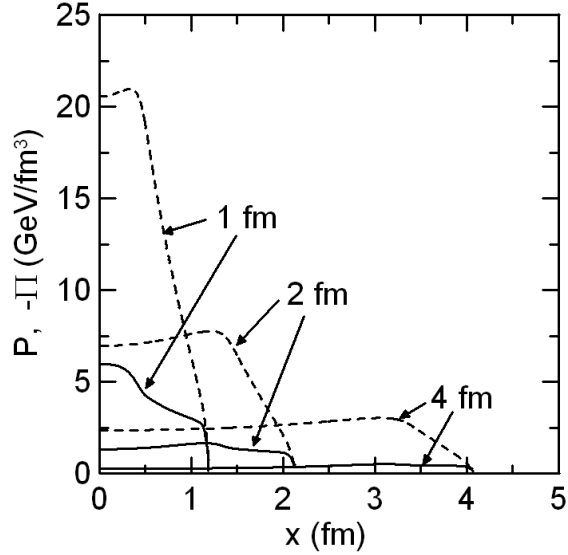}
\caption{The evolution of the pressure (dotted line) and bulk viscosity (solid line)
in NLCDH using $a=0.1$ for $t=1$, $2$ and $3$ fm.}
\label{press_pi}
\end{minipage}
\end{figure}

As was discussed in \cite{dkkm2}, the propagation to vacuum in LCDH, a
stationary wave is formed and the pressure and the bulk viscosity should
satisfy the relation $P=-\Pi$ at the boundary. In Fig. \ref{press_pi}, the
pressure (dotted line) and the bulk viscosity (solid line) of NLCDH are
shown. One can see that the relation $P=-\Pi$ at the boundary is still
satisfied even in NLCDH.

\subsection{Nonperiodic oscillations}

\begin{figure}[ptb]
\begin{minipage}{.45\linewidth}
\includegraphics[scale=0.3]{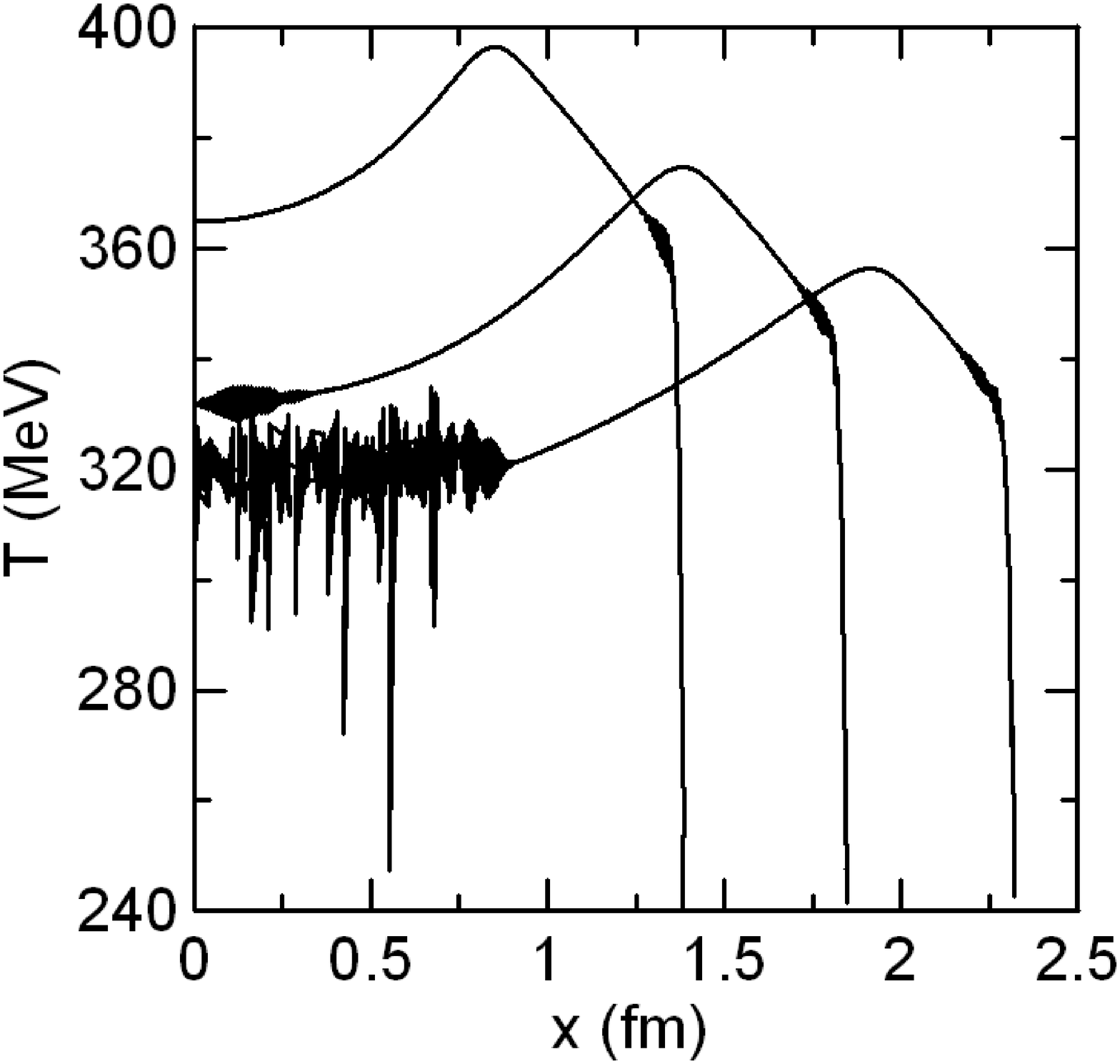}
\caption{The evolution of the temperature with the Landau initial condition
in NLCDH using $a=1$ and $b=6$ for $t=1.44$, $2.04$ and $2.64$ fm from the top.}
\label{tur_temp}
\end{minipage}
\hspace{1cm} 
\begin{minipage}{.45\linewidth}
\includegraphics[scale=0.3]{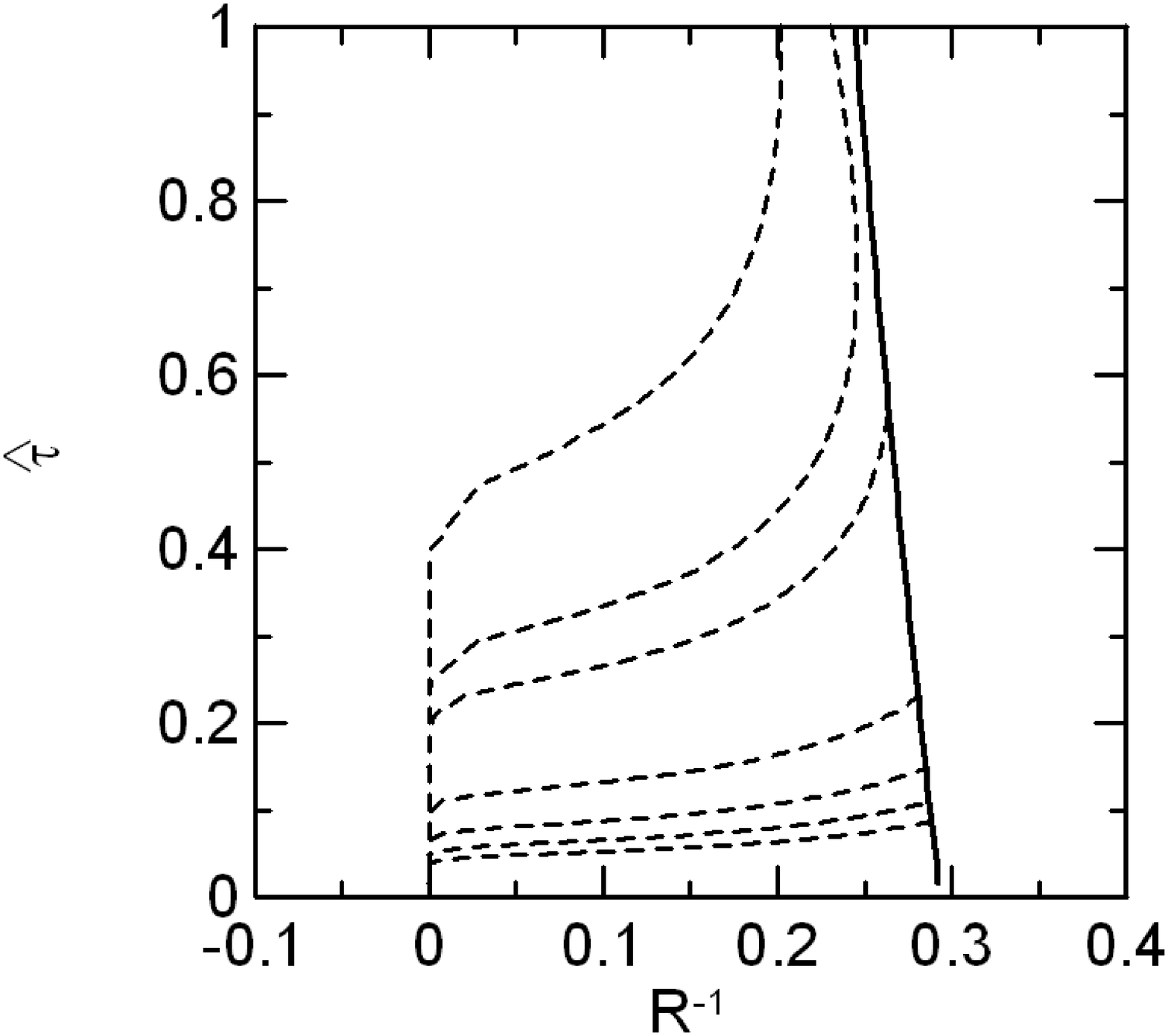}
\caption{The trajectory of the fluid element at the central rapidity region
on the phase diagram as a function of $\hat{\tau}$ and $R_0$
for $a=0.5$, $0.8$, $1$ $2$, $3$, $4$ and $5$ from the top fixing $b=6$.}
\label{phase}
\end{minipage}
\end{figure}

In \cite{dkkm3}, we pointed out that the numerical calculation of LCDH
becomes unstable near the central rapidity region and the nonperiodic
oscillations appear. It is interesting to note that this scenario persists
even in the NLCDH scheme. In Fig.\ref{tur_temp}, we show the evolution of
the temperature calculated in NLCDH with $a=1$, $b=6$ for $t=1.44$, $2.44$
and $2.64$ fm from the top. We use the Landau initial condition with the
initial temperature $590$ MeV and the initial size $0.7$ fm. One can see
that the nonperiodic oscillations evolve with time in the center of the
fluid.

Interestingly, the appearance of nonperiodic oscillations has a regularity.
We investigate the parameter dependence of how these oscillations emerge. In
Fig.\ref{phase}, we plot the trajectories of the fluid element at the
central rapidity region as a function of $\hat{\tau}$ and $R$, for the same
initial condition. Here, $\hat{\tau}$ is the scaled proper time $\hat{\tau}%
=\tau /\tau_{\mathrm{R}}$ and $R$ is the Reynolds number defined as $%
R=-\left( \varepsilon+P\right) /\Pi$ \cite{dkkm3}. The dotted lines show the
trajectories for $a=0.5$, $0.8$, $1$, $2$, $3$, $4$ and $5$ from the top,
respectively. The parameter $b$ is fixed to $6$. We follow the trajectory of
the fluid element in this plane up to the point where the oscillation
emerges for each value of $a$. Thus, the line formed by the endpoints of
these trajectories defines the critical line for the appearance of the
oscillations, indicated by the solid line. One note that the trajectories
without oscillation, $a=0.5$ and $0.8$, do not cross this line. We confirm
numerically that, for various sets of parameters, the oscillations appear
only when a trajectory crosses the critical line. Instabilities have also
been analised for the scaling solution in the framework of the first order
theory \cite{mishust}.

\section{Extended irreversible thermodynamics}

\label{chap:6}

In this paper, we derived our equation by introducing the memory effect and
the finite size effect. As was pointed out, there are several different
approaches to derive the relativistic hydrodynamics consistent with
causality. In this section, we briefly review the derivation based on the
extended irreversible thermodynamics (EIT) \cite{Jou} and show that our
formulation and the extended irreversible thermodynamics give same
conclusion.

The usual thermodynamics describes the thermal equilibrium state which can
be described by the so-called thermodynamic variables; energy, volume and
number of particles. The extended irreversible thermodynamics is the
extension of the usual thermodynamics so as to describe the non-equilibrium
state, which is characterized by not only the thermodynamic variables but
also irreversible currents. Then the first law of thermodynamics in the
local rest frame is 
\begin{eqnarray}
dS=\frac{\partial S}{\partial E}dE+\frac{\partial S}{\partial V}dV+\frac{%
\partial S}{\partial N}dN+\frac{\partial S}{\partial\tilde{\Pi}}d\tilde{\Pi}+%
\frac{\partial S}{\partial\tilde{\pi}^{ij}}d\tilde{\pi}^{ij}+\frac{\partial S%
}{\partial\tilde{\nu}^{i}}d\tilde{\nu}^{i},  \label{eqn:1stlaw1}
\end{eqnarray}
where $\tilde{\Pi}$, $\tilde{\pi}^{ij}$ and $\tilde{\nu}^{i}$ are,
respectively, the bulk viscosity, the shear viscosity and the heat
conduction times $V=1/\sigma$, following the definition of this paper. On
the other hand, the entropy is expanded around the equilibrium state of ($%
E,V,N$) as 
\begin{eqnarray}
S\left( E,V,N,\tilde{\Pi},\tilde{\pi},\tilde{q}\right) -S\left( E,V,N\right)
_{0}=\frac{1}{2}\left. \frac{\partial^{2}S}{\partial\tilde{\Pi }^{2}}%
\right\vert _{0}\tilde{\Pi}^{2}+\frac{1}{2}\left. \frac{\partial^{2}S}{%
\partial\tilde{\pi}^{ij}\partial\tilde{\pi}^{lm}}\right\vert _{0}\tilde {\pi}%
^{ij}\tilde{\pi}^{lm}+\frac{1}{2}\left. \frac{\partial^{2}S}{\partial\tilde{%
\nu}^{i}\partial\tilde{\nu}^{j}}\right\vert _{0}\tilde{\nu }^{i}\tilde{\nu}%
^{j},
\end{eqnarray}
where the suffix $0$ denotes $\left( \tilde{\Pi},\tilde{\pi}^{ij},\tilde{\nu 
}^{i}\right) =0$. It should be noted that we omitted the mixed derivatives
with different tensority like $\partial^{2}S/\partial\tilde{\Pi}\partial 
\tilde{\nu}^{i}$, for simplicity. As a matter of fact, such a term can be
small according to the Curie principle \cite{koide1}. Here, we used that the
first derivatives vanishes because the entropy should be maximum in the
equilibrium, and ignored higher order derivatives. By comparing the two
expressions, we have 
\begin{align}
\frac{\partial S}{\partial\tilde{\Pi}} & =-\beta_{1}\frac{1}{T}\Pi, \\
\frac{\partial S}{\partial\tilde{\pi}^{ij}} & =-\beta_{2}\frac{1}{T}\pi
_{ij}, \\
\frac{\partial S}{\partial\tilde{\nu}^{i}} & =-\beta_{3}\frac{1}{T}{\nu}_{i},
\end{align}
where $\beta_{i}/T$ is a function only of the usual thermodynamic variables, 
$\left( E,V,N\right) $. Inserting these results into the first law (\ref%
{eqn:1stlaw1}), we obtain 
\begin{equation}
TdS=dE+PdV-\mu dN-\beta_{1}\Pi d\tilde{\Pi}-\beta_{2}\pi_{\mu\nu}d\tilde{\pi 
}^{\mu\nu}-\beta_{3}\nu_{\mu}d\tilde{\nu}^{\mu},  \label{eqn:1stlaw2}
\end{equation}
expressed in a covariant form. From the equation of continuity of the
energy-momentum tensor and particle flux, we have 
\begin{align}
\sigma\frac{d}{d\tau}\tilde{\varepsilon}+\left( P+\Pi\right) \partial_{\mu
}u^{\mu}+u_{\mu}\partial_{\nu}\pi^{\mu\nu} & =0,  \label{eqn:aaa} \\
\sigma\frac{d}{d\tau}\tilde{n}+\partial_{\mu}\nu^{\mu} & =0.  \label{eqn:bbb}
\end{align}
By combining Eqs. (\ref{eqn:1stlaw2}), (\ref{eqn:aaa}) and (\ref{eqn:bbb})
together, we have 
\begin{equation}
\partial_{\mu}S^{\mu}=\sigma\frac{d}{d\tau}\tilde{s}-\partial_{\mu}\left( 
\frac{\mu}{T}\nu^{\mu}\right) =Q,
\end{equation}
where 
\begin{equation}
Q=-\frac{\Pi}{T}\left( \beta_{1}\sigma\frac{d\tilde{\Pi}}{d\tau}%
+\partial_{\mu}u^{\mu}\right) -\nu_{\mu}\left( \frac{\beta_{3}}{T}\sigma%
\frac{d\tilde{\nu}^{\mu}}{d\tau}+\partial^{\mu}\frac{\mu}{T}\right) -\frac{%
\pi_{\mu\nu}}{T}\left( \beta_{2}\sigma\frac{d\tilde{\pi}^{\mu\nu}}{d\tau}%
+\partial^{\mu}u^{\nu}\right) ,
\end{equation}
and the entropy four flux is defined by 
\begin{equation}
S^{\mu}=su^{\mu}-\frac{\mu}{T}\nu^{\mu}.
\end{equation}

To satisfy the algebraic positivity of this entropy production, we obtain
the equations of the irreversible current, 
\begin{eqnarray}
\partial_{\mu}u^{\mu}+\beta_{1}\sigma\frac{d\tilde{\Pi}}{d\tau} & = -
\alpha_{1}\Pi, \\
P^{\mu\nu\alpha\beta}\left( \partial_{\alpha}u_{\beta} -\beta_{2}\sigma 
\frac{d\tilde{\pi}_{\alpha\beta}}{d\tau}\right) & = \alpha_{2}\pi^{\mu\nu },
\\
P^{\mu\nu}\left( \partial_{\nu}\frac{\mu}{T} + \frac{\beta_{3}}{T}\sigma 
\frac{d \tilde{\nu}_{\nu}}{d\tau}\right) & = - \alpha_{3}\nu^{\mu},
\end{eqnarray}
where the projection operators are defined by 
\begin{eqnarray}
P^{\mu\nu} & = g^{\mu\nu}-u^{\mu}u^{\nu}, \\
P^{\mu\nu\alpha\beta} & = \frac{1}{2}\left( P^{\mu\alpha}P^{\nu\beta
}+P^{\mu\beta}P^{\nu\alpha}\right) -\frac{1}{D}P^{\mu\nu}P^{\alpha\beta},
\end{eqnarray}
with $D$ is the spatial dimension. Here, $\alpha_{i}$ is a positive
parameter. One can easily see that the equation of, for example, the bulk
viscosity is nothing but the equation obtained in our formulation, by
setting $\zeta= 1/\alpha_{1}$ and $\tau_{\mathrm{R}} = \beta_{1}/\alpha_{1}$.

To our best knowledge, this is the derivation of the relativistic
hydrodynamics based on the extended irreversible thermodynamics for the
first time. As just described, our formulation and the extended irreversible
thermodynamic derives the same equation. The only difference is that the
thermodynamic variables $\varepsilon$ and $P$ in NLCDH satisfy usual
thermodynamic relation, while the variables obey the extended thermodynamic
relation in the hydrodynamic equation of the extended irreversible
thermodynamics.

Note that it is sometimes said that the concept of thermodynamics is
extended in the IS theory. However, as is discussed in Appendix \ref{app:is}
in detail, the first law of thermodynamics is modified in a different way in
the IS theory, without increasing the number of thermodynamical variables. 

\section{Relation between NLCDH and the Israel-Stewart theory}

\label{chap:7}

As was shown in the previous section, the equation obtained in our
formulation can be derived also from the extended irreversible
thermodynamics.

Exactly speaking, the similar non-linear term appears even in the IS theory,
although in the so-called truncated version of IS theory the non-linear
terms are ignored. Then the equation of the bulk viscosity in the original
IS theory is given by 
\begin{equation}
\tau _{\mathrm{R}}\frac{d\Pi }{d\tau }+\Pi =-\zeta \partial _{\mu }u^{\mu }-%
\frac{\tau _{\mathrm{R}}}{2}\Pi \partial _{\mu }u^{\mu }-\frac{\zeta T}{2}%
\Pi \frac{d}{d\tau }\left( \frac{\tau _{\mathrm{R}}}{\zeta T}\right) .
\label{eqn:bulk_is}
\end{equation}%
The last two terms on the r. h. s. are ignored in the truncated IS theory.
For detailed derivation, see Appendix \ref{app:is}. Here we used the
relation $\zeta =1/\alpha _{1}$ and $\tau _{\mathrm{R}}=\beta _{0}/\alpha
_{1}$.

One can easily see that there are two differences between NLCDH and IS
theory. One is the coefficient of the non-linear term $\tau _{\mathrm{R}}\Pi
\partial _{\mu }u^{\mu }$. In NLCDH, the coefficient is given by just $1$,
but it is $1/2$ in the IS theory. The other is the last term of Eq. (\ref%
{eqn:bulk_is}), which does not appear in NLCDH.

\begin{figure}[ptb]
\includegraphics[scale=0.3]{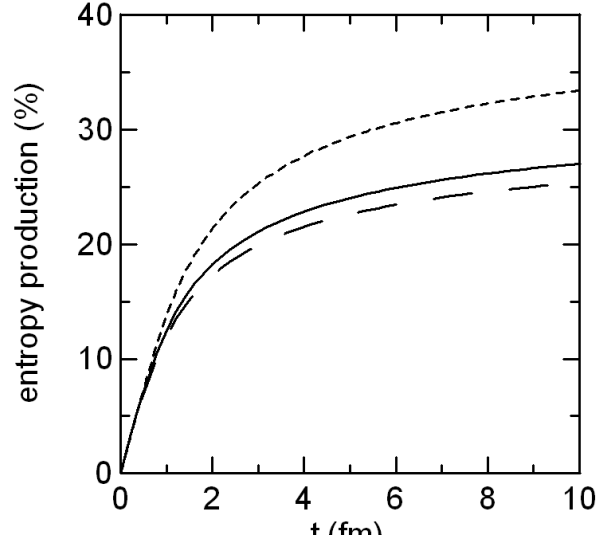} 
\caption{ The evolution of the entropy production with the Landau initial
condition using $a=0.1$. The dotted, solid and dashed lines represent the
calculations of LCDH, NLCDH and the IS theory, respectively.}
\label{entropy}
\end{figure}

As was shown in Sec. \ref{chap:numeri}, the hydrodynamic equation is
stabilized by the non-linear term in NLCDH. In the IS theory, we could not
show the stability of the IS theory analytically because of the last term of
Eq. (\ref{eqn:bulk_is}). However, the numerical simulation shows that the IS
theory is more stable than the truncated IS theory (or LCDH). We calculate
the shock formation and the fluid expansion with the same parameters and
initial conditions, as was discussed in Sec. \ref{sec:shock} and \ref%
{sec:landau}, in the IS theory. We found that the numerical calculations are
stable as in the case of NLCDH and the behaviors of the IS theory is similar
to that of the NLCDH. In Fig \ref{entropy}, we plot the entropy production
as a function of time in the calculation of the expansion to vacuum with the
Landau initial condition. One can see that the entropy production of NLCDH
and the IS theory is smaller than that of LCDH (or the truncated IS theory)
and the IS theory is most close to the Ideal fluid.

The similar reduction due to non-linear terms also appears even for the
shear viscosity as was numerically studied in \cite{heinzson2}. In the bulk
case, we can show explicitly that the suppression happens because of the
minimum value of $\Pi$ guanteed by the nonlinear term. In the case of the
full IS theory, it is not obvious why this happens, but surprizingly the net
effect is very close to ours.

The non-periodic oscillation appears even in the IS theory. It should,
however, be noted that in these simulations, we introduced the additional
viscosity \cite{dkkm2}. If we do not use the additional viscosity, we cannot
implement stable numerical simulations even in the IS theory.

It is important to note that the IS theory also can be derived in our
formulation. So far, we employed the memory effect between $\tilde{\Pi}$ and 
$\tilde{F}$ for simplicity reasons. Suppose we apply the same procedure to $%
\sqrt{\tau _{\mathrm{R}}/\zeta T\sigma }\ \Pi $ and $\zeta \sqrt{\tau _{%
\mathrm{R}}/\zeta T\sigma }\ \partial _{\mu }u^{\mu }$ by using the same
memory function, 
\begin{equation}
\sqrt{\frac{\tau _{\mathrm{R}}}{\zeta T\sigma }}\ \Pi =-\int_{\tau
_{0}}^{\tau }d\tau ^{\prime }G\left( \tau ,\tau ^{\prime }\right) \zeta 
\sqrt{\frac{\tau _{\mathrm{R}}}{\zeta T\sigma }}\partial _{\mu }u^{\mu }.
\label{eqn:ouris1}
\end{equation}%
This leads to the following equation for $\Pi $, 
\begin{equation}
\tau _{\mathrm{R}}\frac{d}{d\tau }\Pi +\frac{1}{2}\tau _{\mathrm{R}}\Pi
\partial _{\mu }u^{\mu }+\frac{1}{2}\zeta \Pi T\frac{d}{d\tau }\frac{\tau _{%
\mathrm{R}}}{\zeta T}=-\Pi -\zeta \partial _{\mu }u^{\mu }.  \label{Full_IS}
\end{equation}%
This equation is nothing but the equation of the bulk viscosity in the IS
theory (See Appendix \ref{app:is}). It can also be derived in the framework
of the\ internal-variable theory, although the concept of thermodynamics
should be extended. See Appendix \ref{app:is} and \ref{app:ivt},
respectively.

\section{Concluding remarks}

\label{chap:8}

\begin{center}
\begin{tabular}{c|ccccc}
&  &  &  &  &  \\ 
macroscopic & $1/\partial_{\mu}u^{\mu}\gg\tau_{\mathrm{R}}$ &  & $%
1/\partial_{\mu}u^{\mu}>\tau_{\mathrm{R}}$ &  & $1/\partial_{\mu}u^{\mu}\sim%
\tau_{\mathrm{R}}$ \\ 
&  &  &  &  &  \\ 
scale & (non-relativistic) & \hspace{1cm} & (relativistic) & \hspace{1cm} & 
(ultra-relativistic) \\ 
&  &  &  &  &  \\ \hline
&  &  &  &  &  \\ 
irreversible & linear response &  & + memory effect &  & + finite size effect
\\ 
&  &  &  &  &  \\ 
currents & ( $J=\eta F$ ) &  & ( $J=\int G\ \eta F$ ) &  & ( $\tilde{J}=\int
G\ \tilde{\eta}\ F$ ) \\ 
&  &  &  &  &  \\ 
&  &  &  &  & 
\end{tabular}
\label{table}
\end{center}

In this paper, we extended the previous derivation of the causal dissipative
hydrodynamics to take into account the finiteness of fluid cells and the
memory effects on the extensive measure of irreversible current inside the
fluid cell. The new equation has a non-linear term which suppresses the
effect of viscosity. Thus the behavior of this fluid is more close to that
of the fluid described by the so-called truncated Israel-Stewart theory,
where there is no finite size effect. More importantly, we found that the
non-linear term is necessary to implement stable numerical simulations for
the ultra-relativistic situations, like high initial velocity, high initial
energy density and so on.

In this study, we found that there are three stages in the structure of
hydrodynamics as is summarized in the above table. When the time scale of
microscopic degrees of freedom are clearly separated from those of the
hydrodynamic variables, we can assume that the irreversible currents $J$ are
immediately produced by the thermodynamic forces $F$, that is, $J=\eta F$.
This is realized in the non-relativistic cases, $1/\partial_{\mu}u^{\mu}\gg%
\tau_{\mathrm{R}}$, because the time scale of constituent particle of the
fluid is much faster than the velocity of the fluid. However, in the
relativistic fluids where $1/\partial_{\mu}u^{\mu}>\tau_{\mathrm{R}}$, the
clear separation of the time scales is not necessarily borne out and we have
to take into account the retardation effect in the formation of the
irreversible currents by introducing memory functions $G$, that is, $J=\int
G\ \eta F$. In the ultra-relativistic limit where the relaxation time is
same order as the scale of inhomogeneity, $\tau_{\mathrm{R}%
}\sim1/\partial_{\mu }u^{\mu}$, it is important to consider the effect of
the finite fluid cell volume $1/\sigma$, because the volume of the fluid
cell changes during the hysteresis, $\tilde{J}=\int G\ \tilde{\eta}\ F$.

We further showed that our formulation and the extended irreversible
thermodynamics lead to the same hydrodynamic equation. The only difference
is that the thermodynamic variables $\varepsilon$ and $P$ in NLCDH satisfy
usual thermodynamic relation, while the variables follows the extended
thermodynamic relation in the hydrodynamic equation in the extended
irreversible thermodynamics. This may indicate us the robustness of our
equation.

\begin{center}
\begin{tabular}{c|c|c}
&  &  \\ 
& IVT (see Appendix \ref{app:ivt}) & EIT \\ 
method & simplest moment method &  \\ 
& memory function method (not extensive) & memory function method (extensive)
\\ 
&  &  \\ 
& $\Downarrow$ & $\Downarrow$ \\ 
&  &  \\ 
derived equation & IS equation & equation of this paper \\ 
&  & 
\end{tabular}
\label{map}
\end{center}

On the other hand, our result is different from that of the IS theory. We
showed that the IS theory also seems to be applicable to the
ultra-relativistic cases, although we could not show the positivity of the
mass explicitly. The quantitative difference between our theory and the IS
theory is very small but the IS theory is more close to the behavior to the
ideal fluid. It is also interesting to mention that the equation of the IS
theory can be derived even in our formulation if we introduce a very
peculiar form of thermodynamical variable to introduce the memory effect.
Although our choice seems more natural and simple from the point of view of
the memory function on thermodynamical forces, we need experimental and
other theoretical supports to decide the appropriate forms of
thermodynamical variables. The way of extending the thermodynamics to
irreversible domain is not unique (See also the discussion in Appendix \ref%
{app:ivt} and Ref. \cite{Jou}). The schematic mapping of different theories
are summarized in Fig. \ref{map}. It should be noted that to derive the IS
equation in the memory function method, we have to break the extensivity of
currents because we employ the memory effect to the quantity proportional to 
$1/\sqrt{\sigma}$ as is shown in Eq. (\ref{eqn:ouris1}).

We are interested in the dynamics of fluid, which cannot be described by the
simple Boltzmann equation. It is not obvious but if the theory is still
applicable to the dynamics of a dilute gas as in the non-relativistic case,
the equation should be justified from the kinetic argument such as the
moment method. As a matter of fact, the problem of the simplest moment
method is known and there are several proposals for the improvement \cite%
{stru}. The derivation of our equaiton from the kinetic point of view is
still an open problem.

\hspace{1cm}

T. Koide acknowledges useful discussions with D. Jou. This work is supported
by FAPERJ and CNPq.

\appendix

\section{Israel-Stewart theory}

\label{app:is}

Similar to the extended irreversible thermodynamics, the equations of
irreversible currents are derived by applying the algebraic positivity of
the entropy production in the IS theory. In the derivation of Israel and
Stewart, they used two equations; one is the definition of the entropy four
flux, 
\begin{equation}
S_{\mathrm{IS}}^{\mu }=s\left( T,\mu \right) u^{\mu }-\frac{\mu }{T}\nu
^{\mu }-Q^{\mu },  \label{eqn:is_s}
\end{equation}%
and the other is the expression of $Q^{\mu }$, which is assumed by the
general quadratic form, 
\begin{equation}
TQ^{\mu }=\frac{1}{2}u^{\mu }\left( \beta _{0}\Pi ^{2}+\beta _{2}\pi _{\mu
\nu }\pi ^{\mu \nu }+\beta _{1}\nu _{\mu }\nu ^{\mu }\right) .
\label{eqn:is_q}
\end{equation}%
Here $s\left( T,\mu \right) $ denotes the entropy density in equilibrium for
given temperature $T$ and chemical potential $\mu .$

By using the algebraic positivity of the entropy production $\partial_{\mu
}S_{\mathrm{IS}}^{\mu}\geq0$ as usual, we obtain 
\begin{align}
\alpha_{1}\Pi & =-\partial_{\mu}u^{\mu}-\frac{\beta_{0}}{2}\Pi\partial
_{\alpha}u^{\alpha}-\beta_{0}\frac{d\Pi}{d\tau}-\frac{T}{2}\Pi\frac{d}{d\tau 
}\left( \frac{\beta_{0}}{T}\right) ,  \label{is1} \\
\alpha_{2}\pi^{\mu\nu} & =P^{\mu\nu\alpha\beta}\left( \partial_{\alpha
}u_{\beta}-\frac{\beta_{2}}{2}\pi_{\alpha\beta}\partial_{\lambda}u^{\lambda
}-\beta_{2}\frac{d\pi_{\alpha\beta}}{d\tau}-\frac{T}{2}\pi_{\alpha\beta}%
\frac{d}{d\tau}\left( \frac{\beta_{2}}{T}\right) \right) ,  \label{is2} \\
\alpha_{3}\nu^{\mu} & =P^{\mu\nu}\left( -\partial_{\nu}\frac{\mu}{T}-\frac{%
\beta_{1}}{2T}\nu_{\nu}\partial_{\alpha}u^{\alpha}-\frac{\beta_{1}}{T}\frac{%
d\nu_{\nu}}{d\tau}-\frac{1}{2}\nu_{\nu}\frac{d}{d\tau}\left( \frac{\beta_{1}%
}{T}\right) \right) .  \label{is3}
\end{align}

The derivation of Israel and Stewart depends on the validity of the
assumption (\ref{eqn:is_s}). To derive this equation, it is usefull to
derive the following relation in the equilibrium, 
\begin{eqnarray}
dS_{\left( 0\right) }^{\mu } &=&d\left( su^{\mu }\right) =\beta _{\nu
}dT_{\left( 0\right) }^{\mu \nu }-\frac{\mu }{T}dN_{\left( 0\right) }^{\mu },
\label{Eq_rel} \\
S_{\left( 0\right) }^{\mu } &=&P\beta ^{\mu }+\beta _{\nu }T_{\left(
0\right) }^{\mu \nu }-\frac{\mu }{T}N_{\left( 0\right) }^{\mu },
\label{euler_cov}
\end{eqnarray}%
here, the subscript $\left( 0\right) $ indicates quantities in equilibrium
and $\beta _{\nu }=u_{\nu }/T.$

The fundamental assumption used by Israel and Stewart is the so called
\textquotedblleft release of variations\textquotedblright , which assumes
that (\ref{Eq_rel}) stays valid for a virtual displacement from a
equilibrium state to an arbitrary neighbouring state, 
\begin{equation}
dS^{\mu }=\beta _{\nu }dT^{\mu \nu }-\frac{\mu }{T}dN^{\mu } + O_2.
\label{eqn:is_assum}
\end{equation}
Here, the last term $O_2$ denotes the contribution from the second order deviation from equilibrium.
This postulate enables us to determine the form of the entropy flux in a
near equilibrium system perturbatively. By addition of (\ref{euler_cov}) and
(\ref{eqn:is_assum}) the the entropy four flux in the IS theory is obtained, 
\begin{equation}
S_{\mathrm{IS}}^{\mu }=P\beta ^{\mu }+T^{\mu \nu }\beta _{\nu }-\frac{\mu }{T%
}N^{\mu }-Q^{\mu },  \label{eqn:sis}
\end{equation}%
where $Q^{\mu }$\ is an undetermined second order term in the deviations $%
T^{\mu \nu }-T_{0}^{\mu \nu }$,\ $N^{\mu }-N_{0}^{\mu }$.

We introduce, again, the energy-momentum tensor and particle flux as
follows; 
\begin{align}
T^{\mu \nu }& =\left( \varepsilon +P+\Pi \right) u^{\mu }u^{\nu }-g^{\mu \nu
}\left( P+\Pi \right) +\pi ^{\mu \nu },  \label{eqn:emt2} \\
N^{\mu }& =nu^{\mu }+\nu ^{\mu }.  \label{eqn:hc2}
\end{align}%
It should be noted that, different from the extended irreversible
thermodynamics, the thermodynamic variables $\varepsilon $, $P$ and $n$
satisfy the usual thermodynamic relations by construction. By using Eqs. (%
\ref{eqn:emt2}) and (\ref{eqn:hc2}), we finally obtain Eq. (\ref{eqn:is_s}).


\section{Internal-variable theory}

\label{app:ivt}

The internal-variable theories (IVT) is another approach to derive the
generalized hydrodynamics \cite{Jou, ciancio}. Similarly to the extended
irreversible thermodynamics, the IVT includes additional variables except
for the usual thermodynamic variables. In this appendix, we use the idea of
the IVT and rederive the IS theory.

First, we assume the following modified first law, 
\begin{equation}
TdS=dE+pdV-\mu dN-TdQ ,  \label{eqn:ivt2}
\end{equation}
where $Q$ is an additional variable. As we will see later, this definition
of the modified first law is different from that in the extended
irreversible thermodynamics. Then, the entropy production is given by 
\begin{align}
\sigma\frac{d\tilde{s}}{d\tau}-\partial_{\mu}\left( \alpha\nu^{\mu}\right) =-%
\frac{\Pi}{T}\partial_{\mu}u^{\mu}+\frac{\pi^{\mu\nu}}{T}\partial_{\mu
}u_{\nu}-\nu^{\mu}\partial_{\mu}\alpha-\sigma\frac{dQ}{d\tau}.
\label{eqn:ivt1}
\end{align}
To obtain the same result as the IS theory, we assume $Q$ as follows, 
\begin{align}
Q=\frac{\sigma}{2T}\left( \beta_{1}\tilde{\Pi}^{2} +\beta_{2}\tilde{\pi}%
_{\mu\nu} \tilde{\pi}^{\mu\nu} +\beta_{3} \tilde{\nu}_{\mu}\tilde{\nu}^{\mu
}\right) .  \label{eqn:ivt3}
\end{align}
Then, for the positivity of the r. h. s. of Eq. (\ref{eqn:ivt1}), we obtain
the IS theory, 
\begin{align}
\left( \partial_{\mu}u^{\mu}+\beta_{1} \frac{d \Pi}{d\tau}+ \frac{\Pi T}{2}%
\frac{d}{d\tau}\left( \frac{\beta_{1}}{T}\right) + \beta_{1}\frac{\Pi}{2}%
\partial_{\mu}u^{\mu}\right) & =-\alpha_{1}\Pi, \\
P^{\mu\nu\alpha\beta}\left( \partial_{\alpha}u_{\beta} - \beta_{2}\frac{d
\pi_{\alpha\beta}}{d\tau}-\frac{\pi_{\alpha\beta}T}{2}\frac{d}{d\tau}\left( 
\frac{\beta_{2}}{T}\right) - \beta_{2}\frac{\pi_{\alpha\beta}}{2}%
\partial_{\mu}u^{\mu}\right) & = \alpha_{2}\pi^{\mu\nu}, \\
P^{\mu\nu}\left( \partial_{\mu}\alpha+ \frac{\beta_{3}}{T}\frac{d \nu_{\mu}}{%
d\tau}+\frac{\nu_{\mu}}{2} \frac{d}{d\tau}\left( \frac{\beta_{3}}{T}\right) +%
\frac{\beta_{3}}{T}\frac{\nu_{\mu}}{2}\partial_{\mu}u^{\mu }\right) & =
-\alpha_{3}\nu^{\nu}.
\end{align}

The main difference between the extended irreversible thermodynamics and the
IVT is the definition of the first law. By substituting Eq. (\ref{eqn:ivt3})
into Eq. (\ref{eqn:ivt2}), we obtain 
\begin{align}
TdS=dE+PdV-\mu dN - \beta_{1} \Pi d\tilde{\Pi} - \beta_{2} \pi_{\mu\nu}d%
\tilde{\pi}^{\mu\nu} - \beta_{3} \nu_{\mu}d\tilde{\nu}^{\mu} - T \tilde{\Pi }%
^{2} d\left( \frac{\sigma\beta_{1}}{2T} \right) - T \tilde{\pi}_{\mu\nu }%
\tilde{\pi}^{\mu\nu} d \left( \frac{\sigma\beta_{2}}{2T} \right) - T \tilde{%
\nu}_{\mu}\tilde{\nu}^{\mu} d \left( \frac{\sigma\beta_{3}}{2T} \right) .
\end{align}
One can easily see that the last three term on the r. h. s. do not exist in
the extended irreversible thermodynamics.


\begin{thebibliography}{99}
\bibitem{GeneralHydro} See for example, Hama Y, Kodama T and Socolowski Jr
O, 2005 Braz.J.Phys. 35:24-51; P. Huovinen and P.V. Ruuskanen, 2006 Ann.
Rev. Nucl. Part. Sci. 56, 163; Jean-Yves Ollitrault, 2008 Euro. J. Phys. 29,
275 and references therein.

\bibitem{1st-a} P. Danielewicz and M. Gyulassy, 1985 Phys. Rev. D 31, 53; D.
Teany, 2003 Phys. Rev. C 68, 034913; P. Van, T. S. Biro, 2008 Eur. Phys. J.
ST 155, 201-212.

\bibitem{1st-b} J. Noronha, G. Torrieri and M. Gyulassy, 2008 Phys. Rev. C
78, 024903; B. M\"{u}ller and J. Ruppert, arXiv:0802.2254.

\bibitem{muro} A. Muronga, Phys. Rev. Lett. 88, 2002 062302 [Erratum ibid.
2002 89, 159901].

\bibitem{2nd} A. Muronga, 2007 Phys. Rev. C 76, 014909; P. Romatschke and U.
Romatschke, 2007 Phys. Rev. Lett. 99, 172301; A. K. Chaudhuri,
arXiv:0801.3180; K. Dusling and D. Teaney, 2008 Phys. Rev. C 77, 034905; R.
S. Bhalerao and S. Gupta, 2008 Phys. Rev. C 77, 014902; A. Dumitru E. M\'{a}%
r and Y. Nara, 2007 Phys. Rev. C 76, 024910; S. Pratt, 2008 Phys. Rev. C 77,
024910; D. Molnar and P. Huovinen, arXiv:0806.1367; E. Moln\'{a}r,
arXiv:0807.0544 and references therein.

\bibitem{heinzson} H. Song and U. W. Heinz, 2008 Phys. Rev. C 77, 064901 .

\bibitem{heinzson2} H. Song and U. W. Heinz, 2008 Phys. Rev. C 78, 024902
(2008).

\bibitem{dkkm1} Koide T, Denicol G S, Mota Ph and Kodama T 2007 Phys. Rev. C
75, 034909.

\bibitem{dkkm2} Denicol G S, Kodama T, Koide T and Mota Ph 2008 Phys. Rev. C
78, 034901.

\bibitem{dkkm3} Denicol G S, Kodama T, Koide T and Mota Ph 2008 J. Phys. G
35 115102.

\bibitem{mishust} Torrieri G and Mishustin I 2008 Phys. Rev. C 78, 021901(R).

\bibitem{tk1} Koide T, 2007 Phys.Rev. E 75, 060103(R).

\bibitem{tk2} Koide T and Kodama T 2008 Phys. Rev. E 78, 051107.

\bibitem{muller} I. M\"uller, 1967 Z. Phys. 198, 329; as a review paper, see
I. M\"uller, 1999 Living Rev. Relativity 2, 1.

\bibitem{II} W. Israel and J. M. Stewart, 1979 Ann. Phys. (N.Y.) 118, 341.

\bibitem{Jou} As a review paper, see D. J, J. Casas-V\'azquez and G. Lebon,
1988 Rep. Prog. Phys. 51, 1105; \textit{ibid} 1999 62, 1035.

\bibitem{carter} B. Carter, 1991 Proc. R. Soc. London, Ser A, 433, 45; as a
review paper, see N. Andersson and G. L. Comer, 2007 Living Rev. Relativity
10, 1.

\bibitem{OG} M. Grmela and H. C. \"Ottinger, 1997 Phys. Rev. E 56, 6620.

\bibitem{baier} R. Baier, P Romatschke, D. T. Son, A. O. Starinets and M. A.
Stephanov, JHEP 0804 (2008) 100.

\bibitem{LL} L. D. Landau and E. M. Lifshitz, \textit{Fluid Mechanics},
(Pergamon; Addison-Wesley, London, U.K.; Reading, U.S.A., 1959).

\bibitem{ref:and} R.P.G. Andrade et al, Phys. Rev. Lett. (to be published),
hep-ph:0805.0018.

\bibitem{SPH} L.B. Lucy, 1977 A. J. 82, 1013 , J.J. Monaghan, 1992 Annu.
Rev. Astron. Astrophys. 30, 543.

\bibitem{koide1} See, for example, T. Koide, 2005 J. Phys. G 31, 1055.

\bibitem{ciancio} V. Ciancio and L. Restuccia, 1990 Physica A 162, 489; V.
Ciancio, L. Restuccia and G. A. Kluitenberg, 1990 J. Non-Equilib. Thermodyn.
15, 157; V. Ciancio and J. Verhas, 1991 J. Non-Equilib. Thermodyn. 16, 57.

\bibitem{stru} See, for example, Struchtrup H and Torrilhon M, 2003 Phys.
Flu. 15, 2668.
\end{thebibliography}
\end{document}